\pdfoutput=1
\documentclass[letterpaper,oneside]{SprMixLaTeX}

\usepackage[sectionbib,longnamesfirst]{natbib}
\bibpunct{(}{)}{;}{a}{}{,}
\renewcommand{\cite}{\citep}

\usepackage{makeidx}         
\usepackage{graphicx,marvosym}        
\usepackage{multicol}        
\usepackage[bottom]{footmisc}

\usepackage{eucal,url}
\usepackage{graphics}

\usepackage{array}
\usepackage{latexsym}
\usepackage{amsmath}
\usepackage{amssymb}
\usepackage{color}
\usepackage{comment}
\usepackage[latin1]{inputenc}

\usepackage{algorithmic}
\usepackage{tabularx}
\usepackage{pbox}

\usepackage{psfrag}

\usepackage{verbatim}
\usepackage{import}

\usepackage{lscape}
\usepackage{subfig}
\usepackage{rotating}

\usepackage{epstopdf}

\usepackage[ruled,algochapter]{algorithm2e}

\usepackage{ifpdf}
\ifpdf
    \usepackage[bookmarks=false,colorlinks=false,citecolor=black,linkcolor=black,breaklinks=true,pdftex,pdfstartview=FitH]{hyperref}
    \hypersetup{
        pdfauthor = {Joseph T. Lizier, Mikhail Prokopenko and Albert Y. Zomaya},
        pdftitle = {A framework for the local information dynamics of distributed computation in complex systems},
        pdfsubject = {Version 3.5},
        pdfkeywords = {information storage, information transfer, information modification, information theory, complex systems, self-organization, intrinsic computation, distributed computation, cellular automata, particles, gliders, domains
},
        pdfcreator = {LaTeX with hyperref package},
        pdfproducer = {pdflatex}}
\fi

\ifx\newtheorem\undefined
  \usepackage{amsthm}
\fi

\newlength{\defsep}
\makeindex             

\begin{document}

\frontmatter


%
%

\mainmatter

\setlength{\fboxsep}{0.3mm}



\newcommand{\figu}[1]{Fig.~\ref{fig:#1}}

\newcommand{\eq}[1]{Eq.~(\ref{eq:#1})}
\newcommand{\eqs}[2]{Eq.~(\ref{eq:#1},~\ref{eq:#2})}
\newcommand{\secRef}[1]{Section~\ref{sec:#1}}
\newcommand{\app}[1]{Appendix~\ref{app:#1}}
\newcommand{\tableRef}[1]{Table~\ref{table:#1}}
\newcommand{\caWidth}{0.48\textwidth}
\newcommand{\caHeight}{0.36\textwidth}



\title{A framework for the local information dynamics of distributed computation in complex systems}
\titlerunning{A framework for local information dynamics}

\author{Joseph T. Lizier$^{1,2}$, Mikhail Prokopenko$^{1}$ and Albert Y. Zomaya$^{2}$}
\authorrunning{Joseph T. Lizier, Mikhail Prokopenko and Albert Y. Zomaya}
\institute{$^{1}$CSIRO Computational Informatics, Locked Bag 17, North Ryde, NSW 1670, Australia\\
$^{2}$School of Information Technologies, The University of Sydney, NSW 2006, Australia}

\maketitle

\vspace{0.7 cm}

\begin{abstract}
The nature of distributed computation has often been described in terms of the component operations of universal computation: \textit{information storage}, \textit{transfer} and \textit{modification}. We review the first complete framework that quantifies each of these individual information dynamics on a local scale within a system, and describes the manner in which they interact to create non-trivial computation where ``the whole is greater than the sum of the parts''.
We describe the application of the framework to cellular automata, a simple yet powerful model of distributed computation.
This is an important application, because the framework is the first to provide quantitative evidence for several important conjectures about distributed computation in cellular automata: that blinkers embody information storage, particles are information transfer agents, and particle collisions are 
information modification events.
The framework is also shown to contrast the computations conducted by several well-known cellular automata, highlighting the importance of information coherence in complex computation.
The results reviewed here provide important quantitative insights into the fundamental nature of distributed computation and the dynamics of complex systems, as well as impetus for the framework to be applied to the analysis and design of other systems.
\end{abstract}

\section{\label{intro}Introduction}

The nature of distributed computation has long been a topic of interest in complex systems science, physics, artificial life and bioinformatics. In particular, emergent complex behavior has often been described from the perspective of computation within the system \cite{mitch98a,mitch98b} and has been postulated to be associated with the capability to support universal computation \cite{lang90,wolf84a,casti91}.

In all of these relevant fields, distributed computation is generally discussed in terms of ``memory'', ``communication'', and ``processing''.
Memory refers to the storage of information by some variable to be used in the future of its time-series process. It has been investigated in coordinated motion in modular robots \cite{pro06b}, in the dynamics of inter-event distribution times \cite{goh08}, and in synchronization between coupled systems \cite{mor07}.
Communication refers to the transfer of information between one variable's time-series process and another; it has been shown to be of relevance in neuroscience \cite{wib11a,lind11a,mar12b} and in other biological systems (e.g. dipole-dipole interaction in microtubules \cite{brown99}, and in signal transduction by calcium ions \cite{pahle08}), social animals (e.g. schooling behavior in fish \cite{couz06}), agent-based systems (e.g. the influence of agents over their environments \cite{kly05b}, and in inducing emergent neural structure \cite{lung06}).
Processing refers to the combination of stored and/or transmitted information into a new form; it has been discussed in particular for biological neural networks and models thereof \cite{kin06a,at92,san02,yam94} (where it has been suggested as a potential biological driver), and also regarding collision-based computing (e.g. \cite{jaku97,adama02}, and including soliton dynamics and collisions \cite{edm93}).

Significantly, these terms correspond to the component operations of Turing universal computation: \textbf{information storage}, \textbf{information transfer} (or transmission) and \textbf{information modification}. Yet despite the obvious importance of these \textbf{information dynamics}, until recently there was no framework for either quantifying them individually or understanding how they interact to give rise to distributed computation.

Here, we review the first complete framework \cite{liz07b,liz08a,liz12a,liz10e,liz10d,liz10a,liz13a} which quantifies each of these information dynamics or component operations of computation within a system, and describes how they inter-relate to produce distributed computation.
We refer to the \textit{dynamics} of information for two key reasons here.
First, this approach describes the composition of information in the \textit{dynamic state update} for the time-series process of each variable within the system, in terms of how information is stored, transferred and modified. This perspective of state updates brings an important connection between information theory and dynamical systems.
Second, the approach focuses on the \textit{dynamics} of these operations on information on  a \textit{local scale} in space and time within the system.
This focus on the local scale is an important one. Several authors have suggested that a complex system is better characterized by studies of its local dynamics than by averaged or overall measures \cite{sha06,han92}, and indeed here we believe that quantifying and understanding distributed computation will necessitate studying the information dynamics and their interplay on a local scale in space and time.
Additionally, we suggest that the quantification of the individual information dynamics of computation provides three \textit{axes of complexity} within which to investigate and classify complex systems, allowing deeper insights into the variety of computation taking place in different systems.

An important focus for discussions on the nature of distributed computation have been cellular automata (CAs) as model systems offering a range of dynamical behavior, including supporting complex computations and the ability to model complex systems in nature \cite{mitch98a}. We review the application of this framework to CAs here because there is very clear \textit{qualitative} observation of emergent structures representing information storage, transfer and modification therein \cite{lang90,mitch98a}.
CAs are a critical proving ground for any theory on the nature of distributed computation: significantly, Von Neumann was known to be a strong believer that ``a general theory of computation in `complex networks of automata' such as cellular automata would be essential both for understanding complex systems in nature and for designing artificial complex systems'' (\citet{mitch98a} describing \citet{von66}).

Information theory provides the logical platform for our investigation, and we begin with a summary of the main information-theoretic concepts required. We provide additional background on the qualitative nature of distributed computation in CAs, highlighting the opportunity which existed for our framework to provide quantitative insights. Subsequently, we consider each component operation of universal computation in turn, and describe how to quantify it locally in a spatiotemporal system. As an application, we review the measurement of each of these information dynamics at every point in space-time in several important CAs.
We show that our framework provided the first complete quantitative evidence for a well-known set of conjectures on the emergent structures dominating distributed computation in CAs: that blinkers provide information storage, particles provide information transfer, and particle collisions facilitate information modification.
Furthermore, we describe the manner in which our results implied that the \textit{coherence} of information may be a defining feature of complex distributed computation.
Our findings are significant because these emergent structures of computation in CAs have known analogues in many physical systems (e.g. solitons and biological pattern formation processes, coherent waves of motion in flocks), and as such this work will contribute to our fundamental understanding of the nature of distributed computation and the dynamics of complex systems.
We finish by briefly reviewing the subsequent application of the framework to various complex systems, including in analyzing flocking behavior and in a computational neuroscience setting.

\section{\label{info}Information-theoretic preliminaries}
Information theory \cite{shan48,cover91,mac03} is an obvious tool for quantifying the information dynamics involved in distributed computation. 
In fact, information theory has already proven to be a useful framework for the design and analysis of complex self-organized systems (e.g. see \cite{pro09}).

We begin by reviewing several necessary information theoretic quantities, including several measures explicitly defined for use with time-series processes.
We also describe \emph{local} information-theoretic quantities - i.e. the manner in which information-theoretic measures can be used to describe the information content associated with single observations.

\subsection{Fundamental quantities}
The fundamental quantity is the Shannon \textbf{entropy}, which represents the uncertainty associated with any measurement $x$ of a random variable $X$ (using units in bits):
\begin{equation}
	H_X = -\sum_{x} p(x) \log_{2}{p(x)}
	\label{eq:entropy}.
\end{equation}

The \textbf{joint entropy} of two (or more) random variables \textit{X} and \textit{Y} is a generalization to quantify the uncertainty of the joint distribution of \textit{X} and \textit{Y}:
\begin{equation}
	H_{X,Y} = -\sum_{x,y} p(x,y) \log_{2}{p(x,y)}
	\label{eq:joint}.
\end{equation}

The \textbf{conditional entropy} of \textit{X} given \textit{Y} is the average uncertainty that remains about \textit{x} when \textit{y} is known:
\begin{equation}
	H_{X \mid Y} = -\sum_{x,y} p(x,y) \log_{2}{p(x \mid y)}
	\label{eq:cond}.
\end{equation}

The \textbf{mutual information} (MI) between \textit{X} and \textit{Y} measures the average reduction in uncertainty about \textit{x} that results from learning the value of \textit{y}, or vice versa:
\begin{eqnarray}
	I_{X;Y} = \sum_{x,y} p(x,y) \log_{2}{\frac{p(x,y)}{p(x)p(y)}}
	\label{eq:mi}. \\
	I_{X;Y} = H_X-H_{X \mid Y} = H_Y-H_{Y \mid X}
	\label{eq:mi2}.
\end{eqnarray}
One can also describe the MI as measuring the information contained in $X$ about $Y$ (or vice versa).

The \textbf{conditional mutual information} between \textit{X} and \textit{Y} given \textit{Z} is the mutual information between \textit{X} and \textit{Y} when \textit{Z} is known:
\begin{eqnarray}
	I_{X;Y \mid Z} & = H_{X \mid Z}-H_{X \mid Y,Z}\\
	& =H_{Y \mid Z}-H_{Y \mid X,Z}
	\label{eq:condmi}.
\end{eqnarray}
Importantly, the conditional MI $I_{X;Y \mid Z}$ can be larger or smaller than the unconditioned $I_{X;Y}$ \cite{mac03}; it is reduced by redundant information held by $Y$ and $Z$ about $X$, and increased by synergy between $Y$ and $Z$ about $X$ (e.g. where $X$ is the result of an exclusive-OR or XOR operation between $Y$ and $Z$).

\subsection{Measures for time-series processes}

Next, we describe several measures which are explicitly defined for time-series processes $X$.

The \textbf{entropy rate} is the limiting value of the rate of change of the joint entropy over \textit{k} consecutive values of a time-series process $X$, (i.e. measurements $\vec{x}^{(k)}_n = \left\{ x_{n-k+1}, \ldots , x_{n-1}, x_n \right\}$, up to and including time step $n$, of the random variable $\vec{X}^{(k)}_n = \left\{ X_{n-k+1}, \ldots , X_{n-1}, X_n \right\}$), as \textit{k} increases \cite{cover91,crutch03}:
\begin{eqnarray}
	H_{\mu X} = \lim_{k \rightarrow \infty}{\frac{H_{\vec{X}^{(k)}_n}}{k}} = \lim_{k \rightarrow \infty}{H_{\mu X}' (k)}
	\label{eq:entrate}, \\
	H_{\mu X}' (k) = \frac{H_{\vec{X}^{(k)}_n}}{k}
	\label{eq:entrateK},
\end{eqnarray}
where the limit exists.
Note that $\vec{X}^{(k)}_n$ is a $k$-dimensional \emph{embedding vector} of the \textit{state} of $X$ \cite{takens81}.
A related definition is given by the limiting value of the conditional entropy of the next value of $X$ (i.e. measurements $x_{n+1}$ of the random variable $X_{n+1}$) given knowledge of the previous $k$ values of \textit{X} (i.e. measurements $\vec{x}^{(k)}_{n}$ of the random variable $\vec{X}^{(k)}_n$):
\begin{eqnarray}
	H_{\mu X} = \lim_{k \rightarrow \infty}{H_{X_{n+1} \mid \vec{X}^{(k)}_n}} = \lim_{k \rightarrow \infty}{H_{\mu X} (k)}
	\label{eq:entratecond}, \\
	H_{\mu X} (k) = H_{\vec{X}^{(k+1)}_{n+1}} - H_{\vec{X}^{(k)}_n}
	\label{eq:entratecondK},
\end{eqnarray}
again, where the limit exists.
This can also be viewed as the uncertainty of the next state $\vec{x}^{(k)}_{n+1}$ given the previous state $\vec{x}^{(k)}_{n}$, since $x_{n+1}$ is the only non-overlapping quantity in $\vec{x}^{(k)}_{n+1}$ which is capable of carrying any conditional entropy.
\citet{cover91} point out that these two quantities correspond to two subtly different notions.
These authors go on to demonstrate 
that for stationary processes $X$, the limits for the two quantities $H'_\mu(X)$ and $H_\mu(X)$ exist (i.e. the average entropy rate converges) and are equal.
For our purposes in considering information dynamics, we are interested in the latter formulation $H_\mu(X)$, since it explicitly describes how one random variable $X_{n+1}$ is related to the previous instances $\vec{X}_{n}^{(k)}$.

\citet{grass86a} first noticed that a slow approach of the entropy rate to its limiting value was a sign of complexity. Formally, \citet{crutch03} use the conditional entropy form of the entropy rate (\ref{eq:entratecond})\footnote{$H_{\mu X} (k)$ here is equivalent to $h_\mu (k+1)$ in \cite{crutch03}.} to observe that at a finite block size \textit{k}, the difference $H_{\mu X} (k) - H_{\mu X}$  represents the information carrying capacity in size $k$-blocks that is due to correlations. The sum over all \textit{k} gives the total amount of structure in the system, quantified as the \textbf{effective measure complexity} or \textbf{excess entropy} (measured in bits):
\begin{equation}
	E_X = \sum^{\infty}_{k=0} \left[ H_{\mu X} (k) - H_{\mu X} \right]
	\label{eq:excess}.
\end{equation}


The excess entropy can also be formulated as the mutual information between the semi-infinite past and semi-infinite future of the system:
\begin{equation}
	E_X = \lim_{k \rightarrow \infty}{I_{\vec{X}^{(k)}_n ; \vec{X}^{(k^+)}_{n+1}}}
	\label{eq:predictiveInfo},
\end{equation}
where $\vec{X}^{(k^+)}_{n+1} = \left\{ X_{n+1}, X_{n+2}, \ldots , X_{n+k} \right\}$ is the random variable (with measurements $\vec{x}^{(k^+)}_{n+1} = \left\{ x_{n+1}, x_{n+2}, \ldots , x_{n+k} \right\}$) referring to the \textit{k} future values of the process $X$ (from time step $n+1$ onwards).
This interpretation is known as the \textit{predictive information} \cite{bialek01}, as it highlights that the excess entropy captures the information in a process' past which is relevant to predicting its future.

\subsection{Local information-theoretic measures}

Finally, we note that the aforementioned information-theoretic quantities are \emph{averages} over all of the observations used to compute the relevant probability distribution functions (PDFs).
One can also write down \emph{local} or pointwise measures for each of these quantities, representing their value for one specific observation or configuration of the variables $(x,y,z)$ being observed. The average of a local quantity over all observations is of course the relevant average information-theoretic measure.

Primarily, the \textbf{Shannon information content} or \textbf{local entropy} of an outcome $x$ of measurement of the variable $X$ is \cite{mac03}: 
\begin{align}
	h(x) = - \log_2{p(x)}
	\label{eq:localEntropy}.
\end{align}
Note that by convention we use lower-case symbols to denote local information-theoretic measures throughout this chapter.
The quantity $h(x)$ is simply the information content attributed to the specific symbol $x$, or the information required to predict or uniquely specify that value.
Less probable outcomes $x$ have higher information content than more probable outcomes, and we have $h(x) \geq 0$.
%
Specifically, the Shannon information content of a given symbol $x$ is the \emph{code-length} for that symbol in an optimal encoding scheme for the measurements $X$, i.e. one that produces the minimal expected code length.\footnote{This ``optimal code-length'' may specify non-integer choices; full discussion of the implications here, practical issues in selecting integer code-lengths, and block-coding optimisations are contained in \cite[Chapter 5]{cover91}.}

Now, note that although the PDF $p(x)$ is \textit{evaluated} for $h(x)$ locally at the given observation $x$, it is \textit{defined} using all of the available (non-local) observations of the variable $X$ which would go into evaluation of the corresponding $H(X)$.
That is to say, we define a certain PDF $p(x)$ from all given measurements of a variable $X$: we can measure local entropies $h(x)$ by evaluating $p(x)$ for a given observation $x$, or we can measure average entropies $H(X)$ from the whole function $p(x)$, and indeed we have $H(X) = \left\langle h(x) \right\rangle$ when the expectation value is taken over $p(x)$.

Similarly, we have the \textbf{local conditional entropy} $h(x \mid y) = - \log_2{p(x \mid y)}$ with $H(X \mid Y) = \left\langle h(x \mid y) \right\rangle$.

Next, the \textbf{local mutual information} \cite{fano61} for a specific observation $(x,y)$ is the information held in common between the specific values $x$ and $y$:
\begin{align}
	i(x;y) & = h(x) - h(x \mid y)
	\label{eq:localMiFromShannon}, \\
		& = \log_2{\frac{p(x \mid y)}{p(x)}}
	\label{eq:localMi}.
\end{align}
The local mutual information is the difference in code lengths between coding the value $x$ in isolation (under the optimal encoding scheme for $X$), or coding the value $x$ given $y$ (under the optimal encoding scheme for $X$ given $Y$).
Similarly, we have the \textbf{local conditional mutual information}:
\begin{align}
	i(x;y \mid z) & = h(x \mid z) - h(x \mid y,z)
	\label{eq:localCondMiFromShannon}, \\
		& = \log_2{\frac{p(x \mid y,z)}{p(x \mid z)}}
	\label{eq:localCondMi}.
\end{align}
Indeed, the form of $i(x;y)$ and $i(x;y \mid z)$ are derived directly from four postulates by \citet[ch. 2]{fano61}: once-differentiability, similar form for conditional MI, additivity (i.e. $i(\left\{y_n,z_n\right\};x_n) = i( y_n ;x_n) + i(z_n; x_n \mid y_n)$), and separation for independent ensembles.
This derivation means that $i(x;y)$ and $i(x;y \mid z)$ are uniquely specified, up to the base of the logarithm.
Of course, we have $I(X;Y) = \left\langle i(x;y) \right\rangle$ and $I(X;Y \mid Z) = \left\langle i(x;y \mid z) \right\rangle$, and like $I(X;Y)$ and $I(X;Y \mid Z)$, the local values are symmetric in $x$ and $y$.

Importantly, $i(x;y)$ may be positive or negative, meaning that one variable can either positively inform us or actually \emph{misinform} us about the other.
An observer is misinformed where, conditioned on the value of $y$ the observed outcome of $x$ was \emph{relatively} unlikely as compared to the unconditioned probability of that outcome (i.e. $p(x \mid y) < p(x)$).
Similarly, $i(x;y \mid z)$ can become negative where $p(x \mid y,z) < p(x \mid z)$.

Applied to time-series data, local measures tell us about the \textit{dynamics} of information in the system, since they vary with the specific observations in time, and local values are known to reveal more details about the system than the averages alone \cite{sha01a,sha06}.

\section{\label{background}Cellular automata}
\subsection{\label{introCAs}Introduction to Cellular Automata}
Cellular automata (CA) are discrete dynamical systems consisting of an array of cells which each synchronously update their discrete value as a function of the values of a fixed number of spatially neighboring cells using a uniform rule. Although the behavior of each individual cell is very simple, the (non-linear) interactions between all cells can lead to very intricate global behavior, meaning CAs have become a classic example of self-organized complex behavior.
Of particular importance, CAs have been used to model real-world spatial dynamical processes, including fluid flow, earthquakes and biological pattern formation \cite{mitch98a}.

The neighborhood of a cell used as inputs to its update rule at each time step is usually some regular configuration. In 1D CAs, this means the same range \textit{r} of cells on each side and including the current value of the updating cell.
One of the simplest variety of CAs -- 1D CAs using binary values, deterministic rules and one neighbor on either side ($r=1$) -- are known as the \textit{Elementary CAs}, or \textit{ECAs}.
Example evolutions of ECAs from random initial conditions may be seen in \figu{54-raw} and \figu{22-raw}. For more complete definitions of CAs, including the definition of the Wolfram rule number convention for specifying update rules, see \citet{wolf02}.

\citet{wolf84a,wolf02} sought to classify the asymptotic behavior of CA rules into four classes: I. Homogeneous state; II. Simple stable or periodic structures; III. Chaotic aperiodic behavior; and IV. Complicated localized structures, some propagating. Much conjecture remains as to whether these classes are quantitatively distinguishable, e.g. see \citet{gray03}, however they do provide an interesting analogy (for discrete-state and time) to our knowledge of dynamical systems, with classes I and II representing ordered behavior, class III representing chaotic behavior, and class IV representing complex behavior and considered as lying between the ordered and chaotic classes.

More importantly though, the approach seeks to characterize complex behavior in terms of emergent structure in CAs, regarding \textit{gliders}, \textit{particles} and \textit{domains}.
Qualitatively, a domain may described as a set of background configurations in a CA, for which any given configuration will update to another such configuration in the set in the absence of any disturbance.
Domains are formally defined within the framework of computational mechanics \cite{han92} as spatial process languages in the CA.
Particles are qualitatively considered to be moving elements of coherent spatiotemporal structure. Gliders are particles which repeat periodically in time while moving spatially (repetitive non-moving structures are known as \textit{blinkers}).
Formally, particles are defined within the framework of computational mechanics as a boundary between two domains \cite{han92}; as such, they can also be termed as domain walls, though this is typically used with reference to aperiodic particles.

These emergent structures are more clearly visible when the CA is filtered in some way.
Early filtering methods were hand-crafted for specific CAs (relying on the user knowing the pattern of background domains) \cite{grass83b,grass89}, while later methods can be automatically applied to any given CA.
These include: $\epsilon$-machines \cite{han92}, input entropy \cite{wue99}, local information \cite{hel04}, and local statistical complexity \cite{sha06}.
All of these filtering techniques produce a \textit{single} filtered view of the structures in the CA: our measures of local information dynamics will present several filtered views of the distributed computation in a CA, separating each operation on information.
The ECA examples analyzed in this chapter are introduced in Section \ref{caExamples}.

\subsection{\label{compCAs}Computation in Cellular Automata}
CAs can be interpreted as undertaking distributed computation: it is clear that ``data represented by initial configurations is processed by time evolution" \cite{wolf84a}.
As such, computation in CAs has been a popular topic for study (see \citet{mitch98a}), with a particular focus in observing or constructing (Turing) universal computation in certain CAs. An ability for universal computation is defined to be where ``suitable initial configurations can specify arbitrary algorithm procedures" in the computing entity, which is capable of ``evaluating any (computable) function" \cite{wolf84a}. \citet{wolf84a,wolf84b} conjectured that all class IV complex CAs were capable of universal computation. He went on to state that prediction in systems exhibiting universal computation is limited to explicit simulation of the system, as opposed to the availability of any simple formula or ``short-cut", drawing parallels to the halting problem for universal Turing machines \cite{wolf84a,wolf84b} which are echoed by \citet{lang90} and \citet{casti91}. (Casti extended the analogy to undecidable statements in formal systems, i.e. G\"odel's Theorem).
The capability for universal computation has been proven for several CA rules, through the design of rules generating elements to (or by identifying elements which) specifically provide the component operations required for universal computation: information storage, transmission and modification. Examples here include most notably the Game of Life \cite{con82} and ECA rule 110 \cite{cook04}; also see \citet{lind90} and discussions by \citet{mitch98a}.

The focus on elements providing information storage, transmission and modification pervades discussion of all types of computation in CAs, e.g. \cite{adama02,jaku01}. \citet{wolf84b} claimed that in class III CAs information propagates over an infinite distance at a (regular) finite speed, while in class IV CAs information propagates at an irregular speed over an infinite range. \citet{lang90} hypothesized that complex behavior in CAs exhibited the three component operations required for universal computation. He suggested that the more chaotic a system becomes the more information transmission increases, and the more ordered a system becomes the more information it stores. Complex behavior was said to occur at a phase transition between these extremes requiring an intermediate level of both information storage and transmission: if information propagates too well, \textit{coherent information decays into noise}. Langton elaborates that transmission of information means that the ``dynamics must provide for the propagation of information in the form of signals over arbitrarily long distances", and suggests that particles in CAs form the basis of these signals. To complete the qualitative identification of the elements of computation in CAs, he also suggested that blinkers formed the basis of information storage, and collisions between propagating (particles) and static structures (blinkers) ``can modify either stored or transmitted information in the support of an overall computation". Rudimentary attempts were made at quantifying the average information transfer (and to some extent information storage), via mutual information (although as discussed later this is a symmetric measure not capturing directional transfer). Recognizing the importance of the emergent structures to computation, several examples exist of attempts to automatically identify CA rules which give rise to particles and gliders, e.g. \cite{wue99,epp02}, suggesting these to be the most interesting and complex CA rules.

Several authors however criticize the aforementioned approaches of attempting to classify CAs in terms of their generic behavior or ``bulk statistical properties", suggesting that the wide range of differing dynamics taking place across the CA makes this problematic \cite{han92,mitch98a}. \citet{gray03} suggests that there there may indeed be classes of CAs capable of more complex computation than universal computation alone. More importantly, \citet{han92} criticize the focus on universal computational ability as drawing away from the ability to identify ``generic computational properties", i.e. a lack of ability for universal computation does not mean a CA is not undertaking any computation at all. Alternatively, these studies suggest that analyzing the rich space-time dynamics \textit{within} the CA is a more appropriate focus.
As such, these and other studies have analyzed the \textit{local} dynamics of intrinsic or other specific computation, focusing on particles facilitating the transfer of information and collisions facilitating the information processing. Noteworthy examples here include: the method of applying filters from the domain of computational mechanics by \citet{han92}; and analysis using such computational mechanics filters of CA rules selected via evolutionary computation to perform classification tasks by \citet{mitch94a,mitch96}.
Related are studies which deeply investigate the nature of particles and their interactions, e.g. particle types and their interaction products identified for particular CAs \cite{mitch96,bocc91,mart06}, and rules established for their interaction products by \citet{hord01}.

Despite such interest, until recently there was no complete framework that locally quantifies the individual information dynamics of distributed computation within CAs or other systems. In this review, we describe how the information dynamics can be locally quantified within the spatiotemporal structure of a CA.
In particular, we describe the dynamics of how information storage and information transfer interact to give rise to information processing. Our approach is not to quantify computation or overall complexity, nor to identify universal computation or determine what is being computed; it is simply intended to quantify the component operations in space-time.

\subsection{\label{caExamples}Examples of distributed computation in CAs}
In this chapter, we review analysis of the computation carried out by several important ECA rules:
\begin{itemize}
	\item Class IV complex rules 110 and 54 \cite{wolf02} (see \figu{110-raw} and \figu{54-raw}), both of which exhibit a number of glider types and collisions. ECA rule 110 is the only proven computationally universal ECA rule \cite{cook04}.
	\item Rules 22 and 30 as representative class III chaotic rules \cite{wolf02} (see rule 22 in \figu{22-raw});
	\item Rules 18 as a class III rule which contains domain walls against a chaotic background domain \cite{wolf84c,han92}.
\end{itemize}
These CAs each carry out an \textit{intrinsic} computation of the evolution to their ultimate attractor and phase on it (see \citet{wue99} for a discussion of attractors and state space in finite-sized CAs).
That is to say, we view the attractor as the end point of an intrinsic computation by the CA -- the dynamics of the transient to the attractor may contain information storage, transfer and modification, while the dynamics on the attractor itself can only contain information storage (since the attractor is either a fixed point or periodic process here).
As such, we are generally only interested in studying computation during the transient dynamics here, as non-trivial computation processes.

We also examine a CA carrying out a ``human-understandable'' computational task. Rule $\phi_{par}$ is a 1D CA with range $r=3$ (the 128-bit Wolfram rule number 0xfeed\-ffde\-c1aa\-eec0\-eef0\-00a0\-e1a0\-20a0) that was evolved by \citet{mitch94a,mitch96} to classify whether the initial CA configuration had a majority of 1's or 0's by reaching a fixed-point configuration of all 1's for the former or all 0's for the latter. This CA rule achieved a success rate above 70\% in its task. An example evolution of this CA can be seen in \figu{phi-raw}. The CA appears to carry out this computation using blinkers and domains for information storage, gliders for information transfer and glider collisions for information modification.
The CA exhibits an initial emergence of domain regions of all 1's or all 0's storing information about local high densities of either value. Where these domains meet, a checkerboard domain propagates slowly (1 cell per time step) in both directions, transferring information of a \textit{soft} uncertainty in this part of the CA. Some ``certainty'' is provided where the glider of the leading edge of a checkerboard encounters a blinker boundary between 0 and 1 domains, which stores information about a \textit{hard} uncertainty in that region of the CA. This results in an information modification event where the domain on the opposite side of the blinker to the incoming checkerboard is concluded to represent the higher density state, and is allowed to propagate over the checkerboard.
This new information transfer associated with local decision of which is the higher density state has evolved to occur at a faster speed (3 cells per time step) than the checkerboard uncertainty; it can overrun checkerboard regions, and in fact collisions of opposing types of this strong propagation give rise to the (hard uncertainty) blinker boundaries in the first place.
The final configuration is therefore the result of this distributed computation.

Quantification of the local information dynamics via these \textit{three axes of complexity} (information storage, transfer and modification) will provide quite detailed insights into the distributed computation carried out in a system. 
In all of these CAs we expect local measures of information storage to highlight blinkers and domain regions, local measures of information transfer to highlight particles (including gliders and domain walls), and local measures of information modification to highlight particle collisions.

This will provide a deeper understanding of computation than single or generic measures of bulk statistical behavior, from which conflict often arises in attempts to provide classification of complex behavior.
In particular, we seek clarification on the long-standing debate regarding the nature of computation in ECA rule 22.
Suggestions that rule 22 is complex include the difficulty in estimating the metric entropy (i.e. temporal entropy rate) for rule 22 by \citet{grass86a}, due to ``complex long-range effects, similar to a critical phenomenon" \cite{grass86b}. This effectively corresponds to an implication that rule 22 contains an infinite amount of memory (see Section \ref{excess}). Also, from an initial condition of only a single ``on" cell, rule 22 forms a pattern known as the ``Sierpinski Gasket'' \cite{wolf02} which exhibits clear fractal structure. Furthermore, rule 22 is a 1D mapping of the 2D Game of Life CA (known to have the capability for universal computation \cite{con82}) and in this sense is referred to as ``life in one dimension'' \cite{mcin90}, and complex structure in the language generated by iterations of rule 22 has been identified by \citet{badii97}. Also, we reported in \cite{liz10d} that we have investigated the $C_1$ complexity measure \cite{laf05} (an enhanced version of the variance of the input entropy \cite{wue99}) for all ECAs, and found rule 22 to clearly exhibit the largest value of this measure (0.78 bits to rule 110's 0.085 bits).
On the other hand, suggestions that rule 22 is not complex include its high sensitivity to initial conditions leading to \citet{wolf02} classifying it as class III chaotic. \citet{guto99} claim this renders it as chaotic despite the subtle long-range effects it displays, further identifying its fast statistical convergence, and exponentially long and thin transients in state space (see \citet{wue99}).
Importantly, no coherent structure (particles, collisions, etc.) is found for typical profiles of rule 22 using a number of known filters for such structure (e.g. local statistical complexity \cite{sha06}): this reflects the paradigm shift to an examination of local dynamics rather than generic, overall or averaged analysis. In our approach, we seek to combine this local viewpoint of the dynamics with a quantitative breakdown of the individual elements of computation, and we will review the application to rule 22 in this light.

\section{\label{storage}Information Storage}
In this section we review the methods to quantify information storage on a local scale in space and time, as presented in \cite{liz12a}. We describe how total information storage used in the future is captured by excess entropy, and introduce active information storage to capture the amount of information storage that is currently in use. We review the application of local profiles of both measures to cellular automata.

\subsection{\label{excess}Excess entropy as total information storage}
Although discussion of information storage or memory in CAs has often focused on periodic structures (particularly in construction of universal Turing machines), information storage does not necessarily entail periodicity. The excess entropy \eqs{excess}{predictiveInfo} more broadly encompasses all types of structure and memory by capturing correlations across all lengths of time, including non-linear correlations. It is quite clear from the predictive information formulation of the excess entropy \eq{predictiveInfo} -- as the information from a process' past that is contained its future -- that it is a measure of the total information storage used in the future of a system.\footnote{In \cite{liz12a} we provide further comment on the relation to the statistical complexity \cite{crutch89}, which measures \emph{all} information stored by the system which \emph{may be used} in the future, while the \emph{excess entropy} measures that information which \emph{is used} by the system \emph{at some point} in the future. The relation between the two concepts is covered in a more general mathematical context by \citet{sha01b}.}

We use the term \textit{univariate excess entropy}\footnote{Called ``single-agent excess entropy'' in \cite{liz12a}.} to refer to measuring the excess entropy for individual variables $X$ using their one-dimensional time-series process, i.e. $E_X = \lim_{k \rightarrow \infty}{I_{\vec{X}^{(k)}_n ; \vec{X}^{(k^+)}_{n+1}}}$ from \eq{predictiveInfo}. This is a measure of the \textit{average} memory for \textit{each variable} $X$.
Furthermore, we use the term \textit{collective excess entropy} to refer to measuring the temporal excess entropy for a collective of variables $\mathbf{X} = \{ X_1, X_2, \ldots, X_m \}$ (e.g. a set of neighboring cells in a CA) using their two-dimensional time-series process. Considered as the mutual information between their joint past and future, i.e. a joint temporal predictive information:
\begin{align}
E_\mathbf{X} = \lim_{k \rightarrow \infty}{I_{\{\vec{X}^{(k)}_{1,n}, \vec{X}^{(k)}_{2,n}, \ldots, \vec{X}^{(k)}_{m,n}\} ; \{\vec{X}^{(k^+)}_{1,n+1}, \vec{X}^{(k^+)}_{2,n+1}, \ldots, \vec{X}^{(k^+)}_{m,n+1}\} }},
\end{align}
this is a measure of the \textit{average} total memory stored in the collective (i.e. stored \textit{collectively} by a set of cells in a CA). Collective excess entropy could be used for example to quantify the ``undiscovered \textit{collective memory} that may present in certain fish schools" \cite{couz06}.

\citet{grass86a,grass86b} studied temporal entropy rate estimates for several ECAs in order to gain insights into their excess entropies.
He revealed divergent collective excess entropy for a number of rules, including rule 22, implying a highly complex process. This case has been described by \citet{lind88} as ``a phenomenon which can occur in more complex environments'', as with strong long-range correlations a semi-infinite sequence ``could store an infinite amount of information about its continuation'' (as per the predictive information form of the excess entropy \eq{predictiveInfo}).
On the other hand, infinite collective excess entropy can also be achieved for systems that only trivially utilise all of their available memory (e.g. simply copying cell values to the right when started from random initial states).
Rule 22 was inferred to have $H_{\mu,N} = 0$ and infinite collective excess entropy, which was interpreted as a process requiring an infinite amount of memory to maintain an aperiodicity \cite{crutch03}.

In attempting to quantify \textit{local} information dynamics of distributed computation here, our focus is on information storage for \textit{single variables or cells} rather than the joint information storage across the collective.
Were the univariate excess entropy found to be divergent (this has not been demonstrated), this may be more significant than for the collective case: divergent collective excess entropy implies that the collective is at least trivially utilizing all of its available memory (and even the chaotic rule 30 exhibits this), whereas divergent univariate excess entropy implies that all cells are individually highly utilizing the resources of the collective in a highly complex process.
Again though, we emphasize that our focus is on local measures in \textit{time} as well as space, which we present in the next section.

\begin{figure*}[t]
	\begin{center}
	\subfloat[Excess Entropy]{\fbox{\label{fig:excessEntropy}\includegraphics[width=0.475\textwidth]{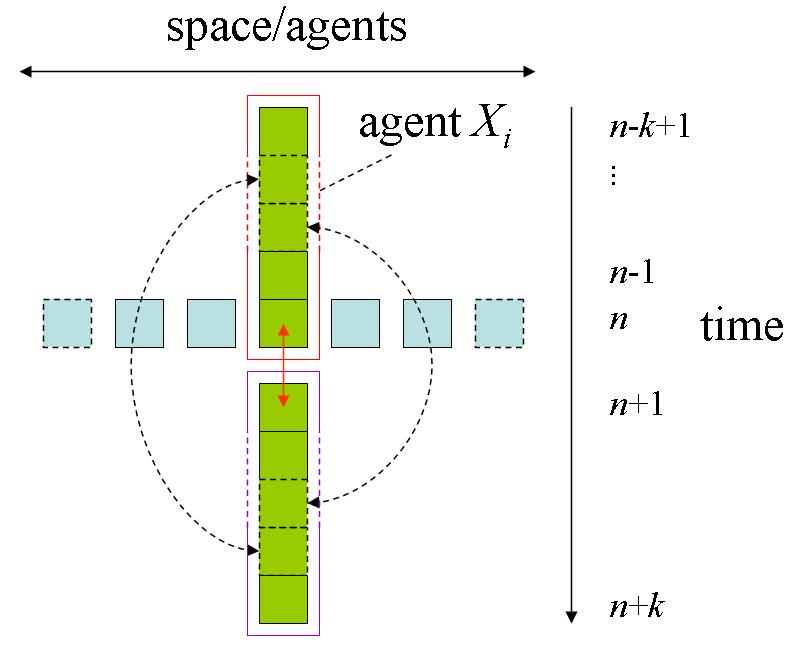}}} \ \ 
	\subfloat[Active Information Storage]{\fbox{\label{fig:activeInfo}\includegraphics[width=0.475\textwidth]{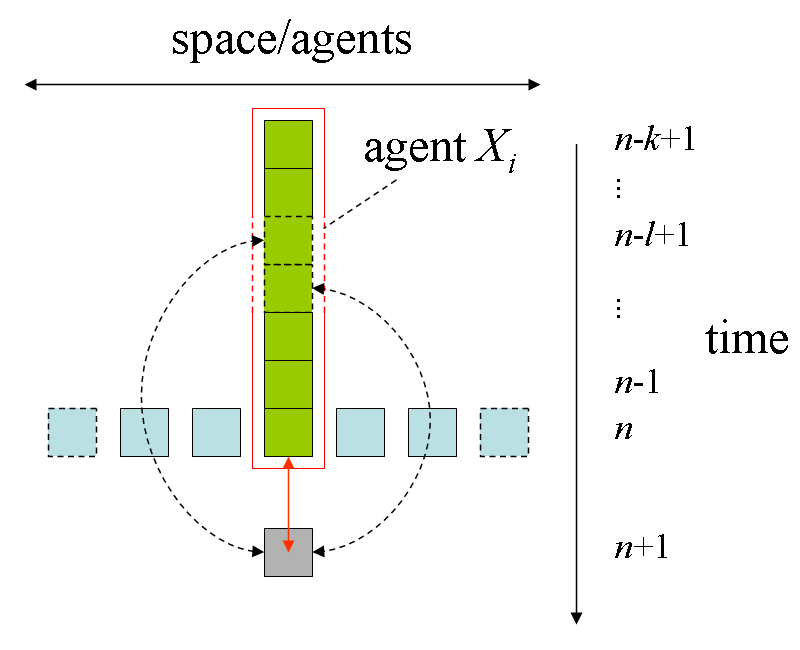}}}
	\end{center}
	\caption{\label{fig:infoStorage} Measures of information storage in the time-series processes of single variables in distributed systems. 
			 \protect \subref{fig:excessEntropy} Excess Entropy: \textit{total} information from the variable's past that is predictive of its future.
			 \protect \subref{fig:activeInfo} Active Information storage: the information storage that is \textit{currently in use} in determining the next value of the variable. The stored information can be conveyed directly through the variable itself or via feedback from neighbors. (NB: This figure is reprinted from \cite{liz12a}, Lizier, J.~T., Prokopenko, M., and Zomaya, A.~Y., {Local measures of information storage in complex distributed computation}, {\em Information Sciences}, 208:39--54, Copyright (2012), with permission from Elsevier.)}
\end{figure*}

First we note that with respect to CAs, where each cell has only a finite number of values $b$ and takes direct influence from only its single past value and the values of a finite number of neighbors, the meaning of (either average or local) information storage being greater $\log_2 b$ bits (let alone infinite) in the time series process of a single cell is not immediately obvious.
Clearly, a cell in an ECA cannot store more than 1 bit of information in isolation.
However, the \textit{bidirectional} communication in CAs effectively allows a cell to store extra information in neighbors (even beyond the immediate neighbors), and to subsequently retrieve that information from those neighbors at a later point in time.
While measurement of the excess entropy does not explicitly look for such \textit{self-influence} communicated through neighbors, it is indeed the method by which a significant portion of information is channeled. Considering the predictive information interpretation in \eq{predictiveInfo}, it is easy to picture self-influence between semi-infinite past and future blocks being conveyed via neighbors (see \figu{excessEntropy}). This is akin to the use of stigmergy (indirect communication through the environment, e.g. see \citet{kly04a}) to communicate with oneself.

A measurement of more than $\log_2 b$ bits stored by a cell on average, or indeed an infinite information storage, is then a perfectly valid result: in an infinite CA, each cell has access to an infinite amount of neighbors in which to store information which can later be used to influence its own future. Note however, that since the storage medium is shared by all cells, one should not think about the total memory as the total number of cells multiplied by this average. The total memory would be properly measured by the collective excess entropy, which takes into account the inherent redundancy here.

Following similar reasoning (i.e. that information may be stored and retrieved from one's neighbors), we note that a variable can store information regardless of whether it is causally connected with itself.
Also, note that a variable can be perceived to store information simply as a result of how that variable is driven \cite{obst13a}, i.e. where information is physically stored elsewhere in the system but recurs in the variable at different time steps (e.g. see the description of information storage in feed-forward loop motifs in \cite{liz12b}).

\subsection{\label{localEE}Local excess entropy}
We now shift focus to local measures of information storage, which have the potential to provide more detailed insights into information storage structures and their involvement in computation than single ensemble measures.

The \textit{local excess entropy} is a measure of how much information a given variable is storing \textit{at a particular point in time} \cite{sha01a}.\footnote{This is as per the original formulation of the local excess entropy by \citet{sha01a}, however this presentation is for a single time-series rather than the light-cone formulation used there.}
The local excess entropy $e_X(n+1)$ of a process is simply the local mutual information \eq{localMi} of the semi-infinite past and future of the process $X$ at the given time step $n+1$:
\begin{equation}
	e_X(n+1) = \lim_{k \rightarrow \infty}{\log_{2}{\frac{p(\vec{x}^{(k)}_n,\vec{x}^{(k^+)}_{n+1})}{p(\vec{x}^{(k)}_n)p(\vec{x}^{(k^+)}_{n+1})}}}
	\label{eq:localExcessEntropy}.
\end{equation}
Note that the excess entropy is the average of the local values, $E_X = \left\langle e_X(n) \right\rangle$.
The limit $k \rightarrow \infty$ is an important part of this definition, since correlations at all time scales should be included in the computation of information storage. Since this is not computationally feasible in general, we retain the notation $e_X(n+1,k)$ to denote finite-$k$ estimates of $e_X(n+1)$.

The notation is generalized for lattice systems (such as CAs) with \textit{spatially-ordered} variables to represent the local excess entropy for cell $X_i$ at time $n+1$ as:
\begin{equation}
	e(i,n+1) = \lim_{k \rightarrow \infty}{\log_{2}{\frac{p(\vec{x}^{(k)}_{i,n},\vec{x}^{(k^+)}_{i,n+1})}{p(\vec{x}^{(k)}_{i,n})p(\vec{x}^{(k^+)}_{i,n+1})}}}
	\label{eq:localExcessEntropyOrdered}.
\end{equation}
Again, $e(i,n+1,k)$ is used to denote finite-$k$ estimates of $e(i,n+1)$. Local excess entropy is defined for every spatiotemporal point $(i,n)$ in the system. (Alternatively, the collective excess entropy can only be localized in time). 

As a local mutual information, the local excess entropy may be positive or negative, meaning the past history of the cell can either positively inform us or actually misinform us about its future. An observer is misinformed where a given semi-infinite past and future are relatively unlikely to be observed together as compared to the product of their marginal probabilities.
Another view is that we have misinformative values when $p(\vec{x}^{(k^+)}_{i,n+1} \mid \vec{x}^{(k)}_{i,n}) < p(\vec{x}^{(k^+)}_{i,n+1})$, meaning that taking the past  $\vec{x}^{(k)}_{i,n}$ into account reduced the probability of the future which was observed $\vec{x}^{(k^+)}_{i,n+1}$.

\subsection{\label{localActiveStorage}Active information storage}
The excess entropy measures the total stored information which will be used \textit{at some point} in the future of the time-series process of a variable, possibly but not necessarily at the next time step $n+1$. In examining the local information dynamics of computation, we are interested in how much of the stored information is actually \textit{in use} at the next time step. As we will see in Section \ref{modification}, this is particularly important in understanding how stored information interacts with information transfer in information processing.
As such, the \textbf{active information storage} $A_X$ was introduced in \cite{liz12a} as the average mutual information between the (semi-infinite) past state of the process and its \textit{next value}, as opposed to its whole (semi-infinite) future:
\begin{equation}
	A_X = \lim_{k \rightarrow \infty}{I(\vec{X}^{(k)}_n;X_{n+1})}.
	\label{eq:activeStorage}
\end{equation}
The \textbf{local active information storage} is then a measure of the amount of information storage in use by the process at a particular time-step $n+1$:
\begin{align}
	a_X(n+1) & = \lim_{k \rightarrow \infty}{\log_{2}{\frac{p(\vec{x}^{(k)}_n,x_{n+1})}{p(\vec{x}^{(k)}_n)p(x_{n+1})}}},
	\label{eq:activeLocalMemory} \\
		& = \lim_{k \rightarrow \infty}{\log_{2}{\frac{p(x_{n+1} \mid \vec{x}^{(k)}_n)}{p(x_{n+1})}}},
	\label{eq:activeLocalMemory2}
\end{align}
and we have $A_X = \left\langle a_X(n) \right\rangle$.
We retain the notation $a_X(n+1,k)$ and $A_X(k)$ for finite-$k$ estimates. Again, we generalize the measure for variable $X_i$ in a lattice system as:
\begin{equation}
	a(i,n+1) = \lim_{k \rightarrow \infty}{\log_{2}{\frac{p(\vec{x}^{(k)}_{i,n},x_{i,n+1})}{p(\vec{x}^{(k)}_{i,n})p(x_{i,n+1})}}},
	\label{eq:activeLocalMemoryCA}
\end{equation}
and use $a(i,n+1,k)$ to denote finite-$k$ estimates there, noting that the local active information storage is defined for every spatiotemporal point $(i,n)$ in the lattice system.

The average active information storage will always be positive (as for the excess entropy), but is bounded above by $\log_2 b$ bits if the variable takes one of $b$ discrete values. The local active information storage is not bound in this manner however, with values larger than $\log_2 b$ indicating that the particular past of an variable provides strong positive information about its next value. Furthermore, the local active information storage can be negative, where the past history of the variable is actually misinformative about its next value. An observer is misinformed where the past history and observed next value are relatively unlikely to occur together as compared to their separate occurrence.


\subsection{\label{localStorageResults}Local information storage results}
In this and subsequent results sections, we review the application of these local measures in \cite{liz07b,liz08a,liz12a,liz10e,liz10d,liz10a,liz13a} to sample CA runs.
As described earlier, we are interested in studying the non-trivial computation during the transient dynamics before an attractor is reached.
Certainly it would be easier to study these information dynamics on attractors -- since the dynamics there are cyclo-stationary (because the attractors in finite-length CAs involve only fixed or periodic dynamics) -- however as described in Section \ref{caExamples} the computation there is trivial.
To investigate the dynamics of the transient, we estimate the required probability distribution functions (PDFs) from CA runs of 10 000 cells, initialized from random values, in order to generate a large ensemble of transient automata dynamics.
We retain only a relatively short 600 time steps for each cell,
in order to avoid attractor dynamics and focus on quasi-stationary transient dynamics during that short time period.
Alternatively, for $\phi_{par}$ we used 30 000 cells with 200 time steps retained. Periodic boundary conditions were used.
Observations taken at every spatiotemporal point in the CA were used in estimating the required PDFs, since the cells in the CA are homogeneous variables and quais-stationarity is assumed over the relatively short time interval.

The results and the figures displayed here were produced using the open source \textit{Java Information Dynamics Toolkit} \cite{liz12d}, which can be used in Matlab/Octave and Python as well as Java.
All results can be reproduced using the Matlab/Octave script \texttt{GsoChapterDemo2013.m} in the \texttt{demos/octave/CellularAutomata} example distributed with this toolkit.
We make estimates of the measures with finite values of $k$, noting that the insights described here could not be attained unless a reasonably large value of $k$ was used in order to capture a large proportion of the correlations. Determination of an appropriate value of $k$ was discussed in \cite{liz12a}, and in \cite{liz08a} for the related transfer entropy measure presented in Section \ref{transfer}. As a rule of thumb, $k$ should at least be larger than the period of any regular background domain in order to capture the information storage underpinning its continuation.

\begin{figure*}
  \begin{center}
		\subfloat[Raw CA]{\label{fig:54-raw}\makebox[\caWidth]{\includegraphics[trim= 0 0 95 0,clip=true,height=\caHeight]{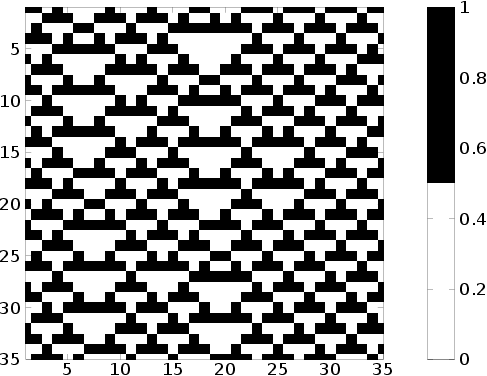}\ \ \ \ \ \ \ \ \ \ \ \ \ }} \ \ 
		\subfloat[$e(i,n,k=8)$]{\label{fig:54-ee-colour-8}\makebox[\caWidth]{\includegraphics[height=\caHeight]{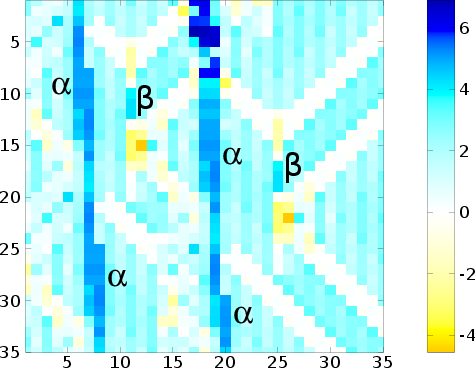}}}

		\subfloat[$a(i,n,k=16)$]{\label{fig:54-active-colour-16}\makebox[\caWidth]{\includegraphics[height=\caHeight]{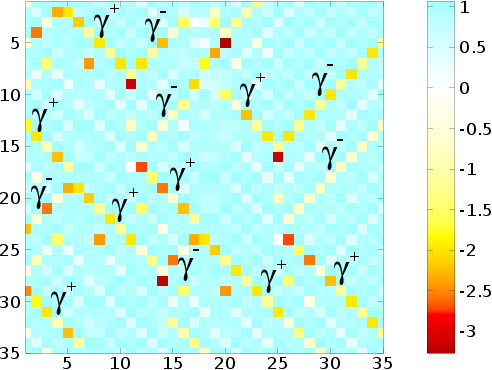}}} \ \ 
		\subfloat[$t(i,j=1,n,k=16)$]{\label{fig:54-te-1-colour-16}\makebox[\caWidth]{\includegraphics[height=\caHeight]{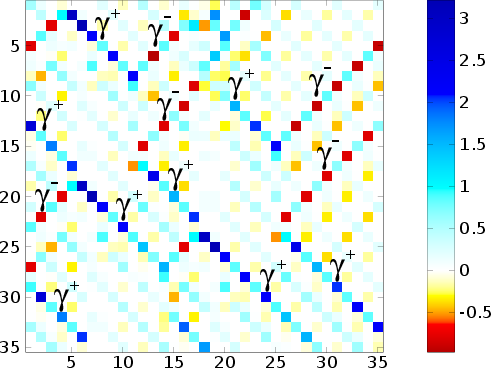}}}
		
		\subfloat[$t(i,j=-1,n,k=16)$]{\label{fig:54-te--1-colour-16}\makebox[\caWidth]{\includegraphics[height=\caHeight]{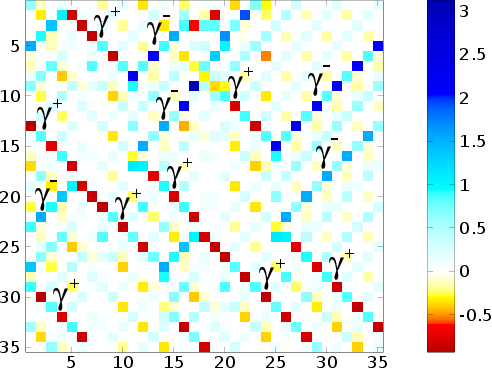}}} \ \ 
		\subfloat[$s(i,n,k=16)$]{\label{fig:54-sep-colour-16}\makebox[\caWidth]{\includegraphics[height=\caHeight]{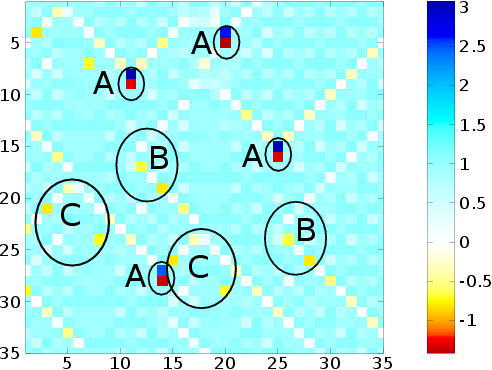}}}
	\end{center}
	\caption[Local information dynamics in ECA rule 54]{\label{fig:54} Local information dynamics in \textbf{rule 54} for the raw values in \subref{fig:54-raw} (black for ``1'', white for ``0''). 35 time steps are displayed for 35 cells, and time increases down the page for all CA plots. All units are in bits.
		\subref{fig:54-ee-colour-8} Local excess entropy $e(i,n,k=8)$; 
		\subref{fig:54-active-colour-16} Local active information storage $a(i,n,k=16)$; 
		Local apparent transfer entropy: \subref{fig:54-te-1-colour-16} one cell to the right $t(i,j=1,n,k=16)$, 
		\subref{fig:54-te--1-colour-16} one cell to the left $t(i,j=-1,n,k=16)$; 
		\subref{fig:54-sep-colour-16} Local separable information $s(i,n,k=16)$. 
	}
\end{figure*}

\begin{figure*}
  \begin{center}
		\subfloat[Raw CA]{\label{fig:110-raw}\makebox[\caWidth]{\includegraphics[trim= 0 0 95 0,clip=true,height=\caHeight]{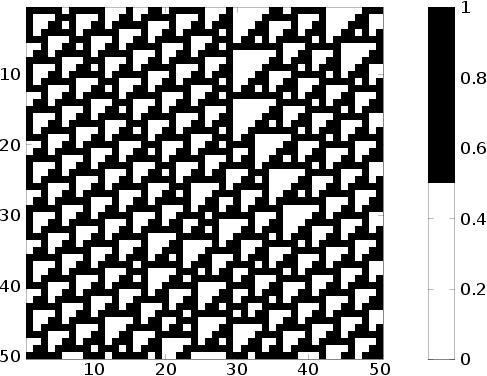}\ \ \ \ \ \ \ \ \ \ \ \ \ }} \ \ 
		\subfloat[$e(i,n,k=8)$]{\label{fig:110-ee-colour-8}\makebox[\caWidth]{\includegraphics[height=\caHeight]{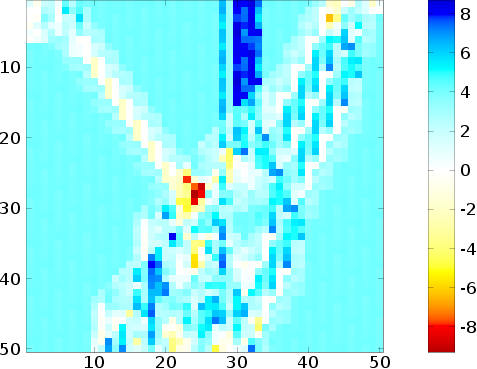}}}

		\subfloat[$a(i,n,k=16)$]{\label{fig:110-active-colour-16}\makebox[\caWidth]{\includegraphics[height=\caHeight]{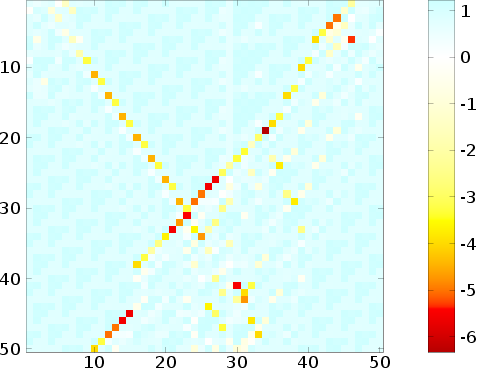}}} \ \ 
		\subfloat[$h_\mu(i,n,k=16)$]{\label{fig:110-entRate-colour-16}\makebox[\caWidth]{\includegraphics[height=\caHeight]{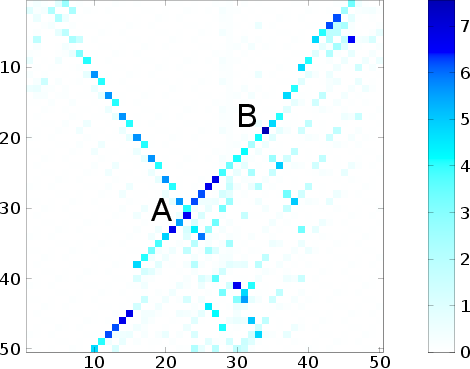}}}
		
		\subfloat[$t(i,j=-1,n,k=16)$]{\label{fig:110-te--1-colour-16}\makebox[\caWidth]{\includegraphics[height=\caHeight]{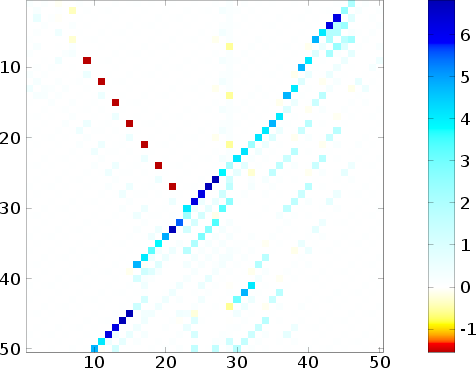}}} \ \ 
		\subfloat[$s(i,n,k=16)$]{\label{fig:110-sep-colour-16}\makebox[\caWidth]{\includegraphics[height=\caHeight]{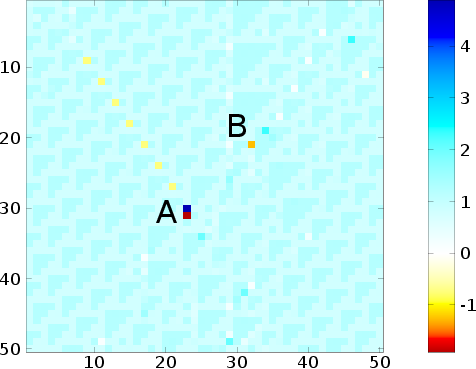}}}
	\end{center}
	\caption[Local information dynamics in ECA rule 110]{\label{fig:110} Local information dynamics in \textbf{rule 110} for the raw values displayed in \subref{fig:110-raw} (black for ``1'', white for ``0''). 50 time steps are displayed for 50 cells, and all units are in bits.
		\subref{fig:110-ee-colour-8} Local excess entropy $e(i,n,k=8)$; 
		\subref{fig:110-active-colour-16} Local active information storage $a(i,n,k=16)$; 
		\subref{fig:110-entRate-colour-16} Local temporal entropy rate $h_\mu(i,n,k=16)$; 
		\subref{fig:110-te--1-colour-16} Local apparent transfer entropy one cell to the left $t(i,j=-1,n,k=16)$; 
		\subref{fig:110-sep-colour-16} Local separable information $s(i,n,k=16)$. 
	}
\end{figure*}

We begin by examining the results for rules 54 and 110, which contain regular gliders against periodic background domains. For the CA runs described above, sample areas of the large CAs are shown in \figu{54-raw} and \figu{110-raw}, while the corresponding local profiles of $e(i,n,k = 8)$ generated are displayed in \figu{54-ee-colour-8} and \figu{110-ee-colour-8}, and the local profiles of $a(i,n,k = 16)$ in \figu{54-active-colour-16} and \figu{110-active-colour-16}.
It is quite clear that positive information storage is concentrated in the vertical gliders or blinkers, and the domain regions.
As expected, these results provide quantitative evidence that the \textbf{blinkers are the dominant information storage entities}.
That the \textbf{domain regions contain significant information storage} should not be surprising, since as a periodic sequence its past does indeed store information about its future.

In fact, the local values for each measure form spatially and temporally periodic patterns in the domains, corresponding to the spatial and temporal periodicities exhibited in the underlying raw values.
Certainly if the dynamics are only composed of a consistent domain pattern (which is deterministic when viewing single cells' time series), then for $a(i,n,k)$ for example we will always have $p(x_{n+1} \mid \vec{x}^{(k)}_n) = 1$ and if $p(x_{n+1})$ is balanced then $a(i,n,k)$ would be constant across the CA.
However, the existence of discontinuities in the domain, e.g. gliders, reduces $p(x_{n+1} \mid \vec{x}^{(k)}_n)$ here, and does so differently for each $\vec{x}^{(k)}_n$ configuration in the domain.
Imbalances in $p(x_{n+1})$ can also contribute to differences in storage across the domain.
These factors leads to the spatiotemporal periodicities of information storage that are observed in the domains.

While the local active information storage indicates a similar amount of stored information in use to compute each space-time point in both the domain and blinker areas, the local excess entropy reveals a larger \textit{total} amount of information is stored in the blinkers. For the blinkers known as $\alpha$ and $\beta$ in rule 54 \cite{hord01} this is because the temporal sequences of the center columns of the blinkers (0-0-0-1, with $e(i,n,k = 8)$ in the range 5.01 to 5.32 bits) are more complex than those in the domain (0-0-1-1 and 0-1, with $e(i,n,k = 8)$ in the range 1.94 to 3.22 bits), even where they are of the same period. We have $e(i,n,k = 8) > 1$ bit here due to the distributed information storage supported by bidirectional communication (as discussed earlier). Such bidirectional communication is also critical to these periodic domain sequences being longer than two time steps -- the maximum period that a binary cell could sustain in isolation (e.g. the period-7 domain in rule 110).

Another area of strong information storage appears to be the ``wake'' of the more complex gliders in rule 110 (see the glider at top right of \figu{110-ee-colour-8} and \figu{110-active-colour-16}). This result aligns well with our observation \cite{liz08a} that the dynamics following the leading edge of regular gliders consists largely of ``non-traveling'' information. The presence of the information storage is shown by both measures, although the relative strength of the total information storage is again revealed only by the local excess entropy.

\begin{figure}[t]
	\begin{center}
		\fbox{\includegraphics[width=0.35\textwidth]{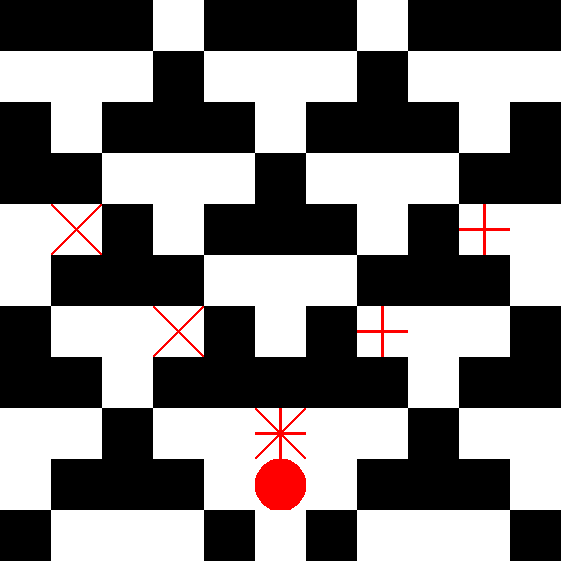}}
	\end{center}
	\caption{\label{fig:54-closeup} Close up of raw values of rule 54. ``x'' and ``+'' mark some positions in the $\gamma^+$ and $\gamma^-$ gliders respectively. Note their point of coincidence in collision type ``A", with ``$\bullet$'' marking the subsequent non-trivial information modification as detected using $s(i,n,k=16)<0$. (Reprinted with permission from \cite{liz10e} J.~T. Lizier, M.~Prokopenko, and A.~Y. Zomaya, ``Information modification and particle collisions in distributed computation,'' \emph{Chaos},  vol.~20, no.~3, p. 037109, 2010. Copyright 2010, AIP Publishing LLC.)}
\end{figure}

Negative values of $a(i,n,k = 16)$ for rules 54 and 110 are also visible in \figu{54-active-colour-16} and \figu{110-active-colour-16}. Interestingly, negative local components of local active information storage measure are concentrated in the traveling glider areas (e.g. $\gamma^{+}$ and $\gamma^{-}$ for rule 54 \cite{hord01}), providing a good spatiotemporal filter of these structures.
This is because when a traveling glider is encountered at a given cell, the past history of that cell (being part of the background domain) is misinformative about the next value, since the domain sequence was more likely to continue than be interrupted.
For example, see the marked positions of the $\gamma$ gliders in \figu{54-closeup}. There we have $p(x_{n+1} \mid \vec{x}^{(k=16)}_n) = 0.25$ and $p(x_{n+1}) = 0.52$: since the next value occurs relatively infrequently after the given history, we have a misinformative $a(n,k=16) = -1.09$ bits. This is juxtaposed with the points four time steps before those marked ``x", which have the same history $\vec{x}^{(k=16)}_n$ but are part of the domain, with $p(x_{n+1} \mid \vec{x}^{(k=16)}_n) = 0.75$ and $p(x_{n+1}) = 0.48$ giving $a(n,k=16) = 0.66$ bits, quantifying the positive information storage there. Note that the points with misinformative information storage are not necessarily those selected by other filtering techniques as part of the gliders: e.g. the finite state transducers technique (using left to right scanning by convention) by \citet{han97} would identify points 3 cells to the right of those marked ``x" as part of the $\gamma^+$ glider.

The local excess entropy produced some negative values around traveling gliders, though these were far less localized on the gliders themselves and less consistent in occurrence than for the local active information storage. This is because the local excess entropy, as measure of total information storage into the future, is more loosely tied to the dynamics at the given spatiotemporal point. The effect of a glider encounter on $e(i,n,k)$ is smeared out in time, and in fact the dynamics may store more positive information in total than the misinformation encountered at the specific location of the glider. For example, glider pairs were observed in \cite{liz12a} to have positive total information storage, since a glider encounter becomes much more likely in the wake of a previous glider. 

\begin{figure*}
  \begin{center}
		\subfloat[Raw CA]{\label{fig:phi-raw}\makebox[\caWidth]{\includegraphics[trim= 0 0 95 0,clip=true,height=\caHeight]{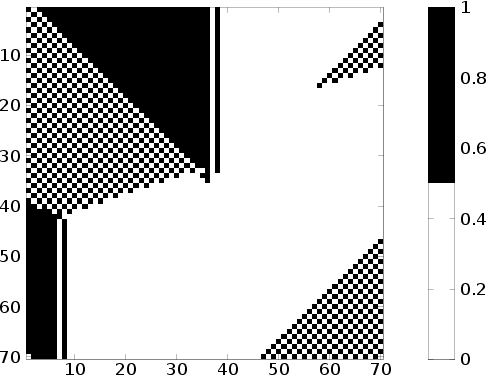}\ \ \ \ \ \ \ \ \ \ \ \ \ }} \ \ 
		\subfloat[$a(i,n,k=10)$]{\label{fig:phi-active-colour-10}\makebox[\caWidth]{\includegraphics[height=\caHeight]{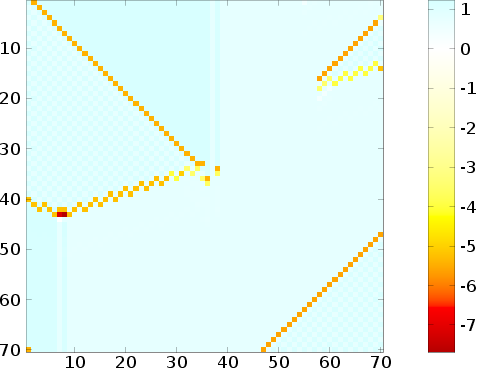}}}
		
		\subfloat[$t(i,j=-1,n,k=10)$]{\label{fig:phi-te--1-colour-10}\makebox[\caWidth]{\includegraphics[height=\caHeight]{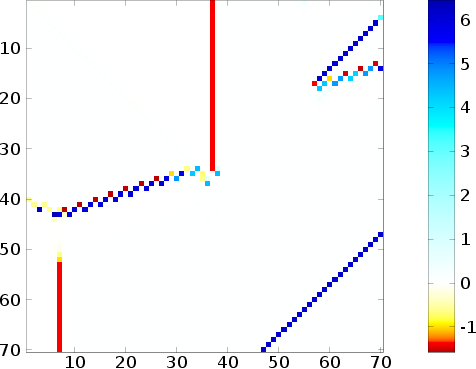}}} \ \ 
		\subfloat[$t^c(i,j=-1,n,k=10)$]{\label{fig:phi-teComp--1-colour-10}\makebox[\caWidth]{\includegraphics[height=\caHeight]{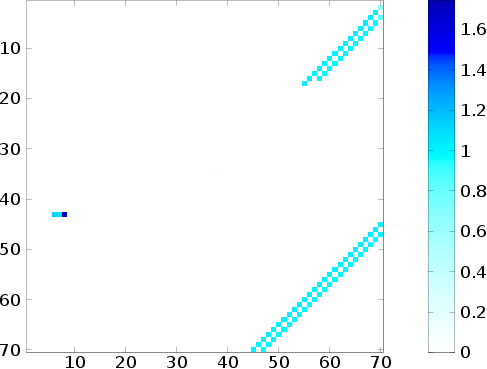}}}
		
		\subfloat[$t(i,j=-3,n,k=10)$]{\label{fig:phi-te--3-colour-10}\makebox[\caWidth]{\includegraphics[height=\caHeight]{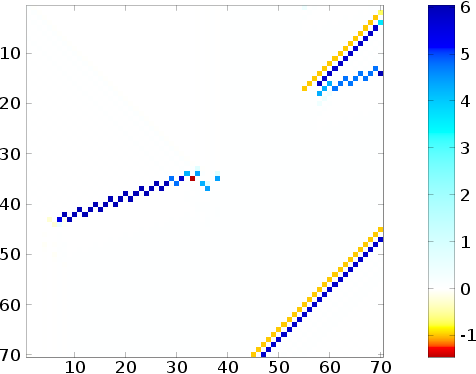}}} \ \ 
		\subfloat[$s(i,n,k=10)$]{\label{fig:phi-sep-colour-10}\makebox[\caWidth]{\includegraphics[height=\caHeight]{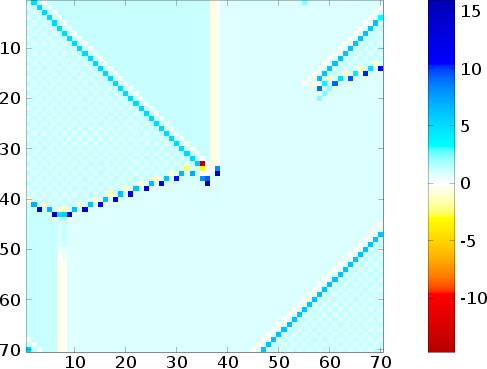}}}
	\end{center}
	\caption[Local information dynamics in $r=3$ rule $\phi_{par}$]{\label{fig:phi} Local information dynamics in $r=3$ \textbf{rule} $\mathbf{\phi_{par}}$ for the raw values displayed in \subref{fig:phi-raw} (black for ``1'', white for ``0''). 70 time steps are displayed for 70 cells, and all units are in bits.
		\subref{fig:phi-active-colour-10} Local active information storage $a(i,n,k=10)$; 
		Local apparent transfer entropy: \subref{fig:phi-te--1-colour-10} one cell to the left $t(i,j=-1,n,k=10)$, 
		and \subref{fig:phi-te--3-colour-10} three cells to the left $t(i,j=-3,n,k=10)$;
		\subref{fig:phi-teComp--1-colour-10} Local complete transfer entropy one cell to the left $t^c(i,j=-1,n,k=10)$;
		\subref{fig:phi-sep-colour-10} Local separable information $s(i,n,k=10)$. 
	}
\end{figure*}

As another rule containing regular gliders against a periodic background domain, analysis of the raw values of $\phi_{par}$ in \figu{phi-raw} provides similar results for $e(i,n,k = 5)$ (not shown, see \cite{liz13a}) and $a(i,n,k = 10)$ in \figu{phi-active-colour-10} here.
One distinction is that the blinker here contains no more stored information than the domain, since it is no more complicated. Importantly, we confirm the information storage capability of the blinkers and domains in this human understandable computation.

Another interesting example is provided by ECA rule 18, which contains domain walls against a seemingly irregular background domain.
We measured the local information profiles for $e(i,n,k=8)$ and $a(i,n,k=16)$ in \cite{liz12a} (shown in that paper, but not here).
Importantly, the most significant negative components of the local active information storage are concentrated on the domain walls: analogous to the regular gliders of rule 54, when a domain wall is encountered the past history of the cell becomes misinformative about its next value. 
There is also interesting information storage dynamics in the background domain for rule 18, discussed in detail in \cite{liz12a}.

\begin{figure*}
  \begin{center}
		\subfloat[Raw CA]{\label{fig:22-raw}\makebox[\caWidth]{\includegraphics[trim= 0 0 95 0,clip=true,height=\caHeight]{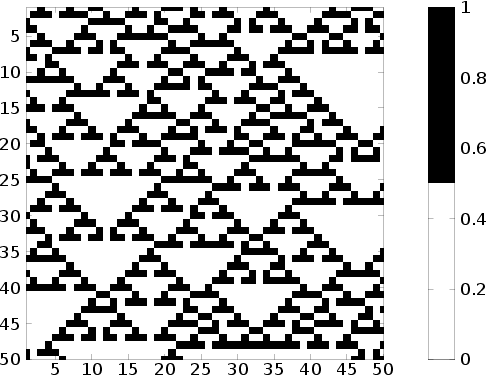}\ \ \ \ \ \ \ \ \ \ \ \ \ }} \ \ 
		\subfloat[$e(i,n,k=8)$]{\label{fig:22-ee-colour-8}\makebox[\caWidth]{\includegraphics[height=\caHeight]{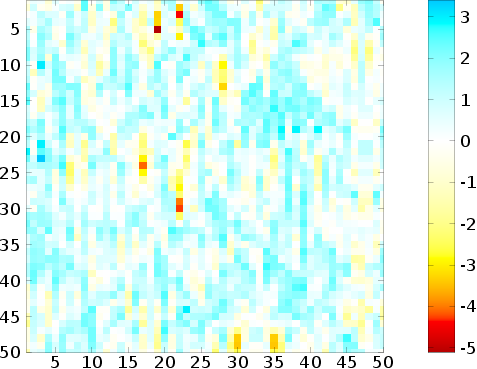}}}
		
		\subfloat[$a(i,n,k=16)$]{\label{fig:22-active-colour-16}\makebox[\caWidth]{\includegraphics[height=\caHeight]{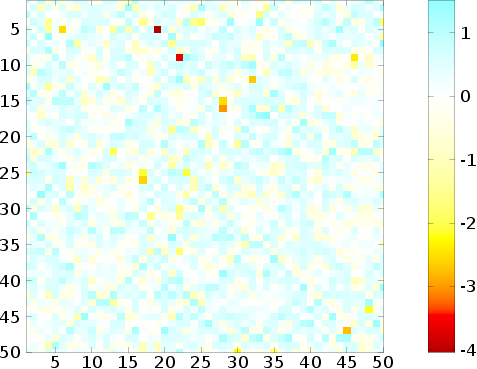}}} \ \ 
		\subfloat[$h_\mu(i,n,k=16)$]{\label{fig:22-entRate-colour-16}\makebox[\caWidth]{\includegraphics[height=\caHeight]{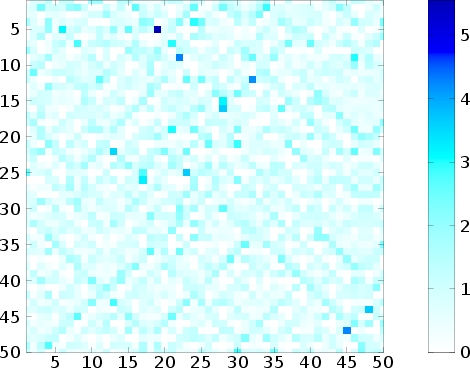}}}
		
		\subfloat[$t(i,j=1,n,k=16)$]{\label{fig:22-te-1-colour-16}\makebox[\caWidth]{\includegraphics[height=\caHeight]{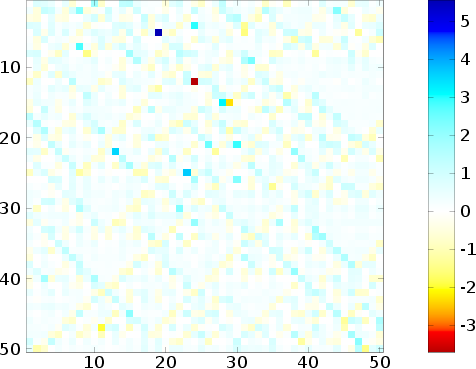}}} \ \ 
		\subfloat[$s(i,n,k=16)$]{\label{fig:22-sep-colour-16}\makebox[\caWidth]{\includegraphics[height=\caHeight]{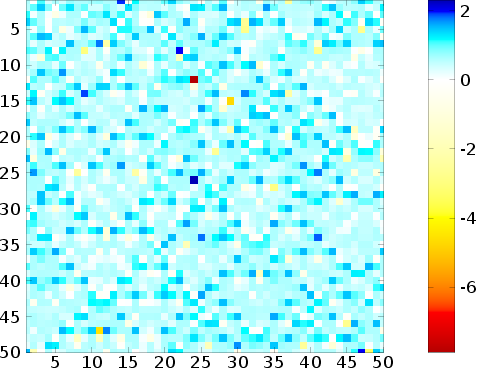}}}
	\end{center}
	\caption[Local information dynamics in ECA rule 22]{\label{fig:22} Local information dynamics in \textbf{rule 22} for the raw values in \subref{fig:22-raw} (black for ``1'', white for ``0''). 50 time steps displayed for 50 cells, and all units are in bits.
		\subref{fig:22-ee-colour-8} Local excess entropy $e(i,n,k=8)$; 
		\subref{fig:22-active-colour-16} Local active information storage $a(i,n,k=16)$; 
		\subref{fig:22-entRate-colour-16} Local temporal entropy rate $h_\mu(i,n,k=16)$; 
		\subref{fig:22-te-1-colour-16} Local apparent transfer entropy one cell to the right $t(i,j=1,n,k=16)$; 
		\subref{fig:22-sep-colour-16} Local separable information $s(i,n,k=16)$. 
		}
\end{figure*}

Finally, we examine ECA rule 22, suggested to have infinite collective excess entropy \cite{grass86a,grass86b} but without any known coherent structural elements \cite{sha06}. For the raw values of rule 22 displayed in \figu{22-raw}, the calculated local excess entropy profile is shown in \figu{22-ee-colour-8}, and the local active information storage profile in \figu{22-active-colour-16}. While information storage certainly occurs for rule 22, these plots provide evidence that there is no coherent structure to this storage. This is another clear example of the utility of examining local information dynamics over ensemble estimates, given the earlier discussion on collective excess entropy for rule 22.

In summary, we have demonstrated that the local active information storage and local excess entropy provide insights into information storage dynamics that, while often similar in general, are sometimes subtly different. While both measures provide useful insights, the local active information storage is the most useful in a real-time sense, since calculation of the local excess entropy requires knowledge of the dynamics an arbitrary distance into the future.\footnote{Calculation of $e(i,n,k)$ using local block entropies analogous to \eq{excess} would also require block entropies to be taken into the future to compute the same local information storage values. Without taking account of the dynamics into the future, we will not measure the information storage that \textit{will} be used in the future of the process, but the information storage that is \textit{likely} to be used in the future.}
Furthermore, it also provides the most specifically localized insights, including filtering moving elements of coherent spatiotemporal structure.
This being said, it is not capable of identifying the information source of these structures; for this, we turn our attention to a specific measure of information transfer.

\section{\label{transfer}Information Transfer}
Information transfer refers to a directional signal or communication of dynamic information from a \textit{source} to a \textit{destination}.
In this section, we review descriptions of how to measure information transfer in complex systems from \cite{liz08a,liz10e,liz13a}, and the associated application to several ECA rules.

\subsection{\label{localTeTheory}Local transfer entropy}

\begin{figure*}[t]
	\subfloat[Transfer Entropy]{\fbox{\label{fig:transferEntropy}\includegraphics[width=0.48\textwidth]{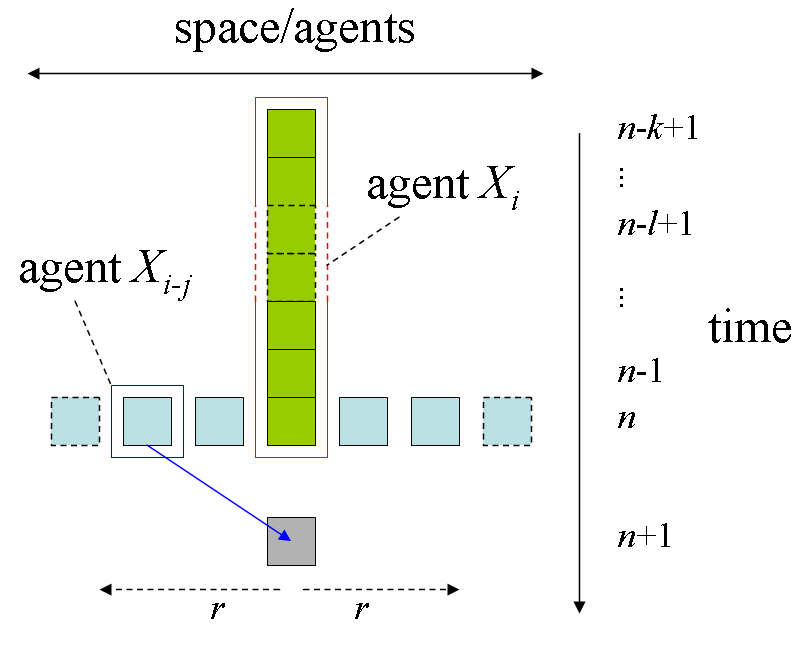}}} \ \ 
	\subfloat[Separable Information]{\fbox{\label{fig:separableInfo}\includegraphics[width=0.48\textwidth]{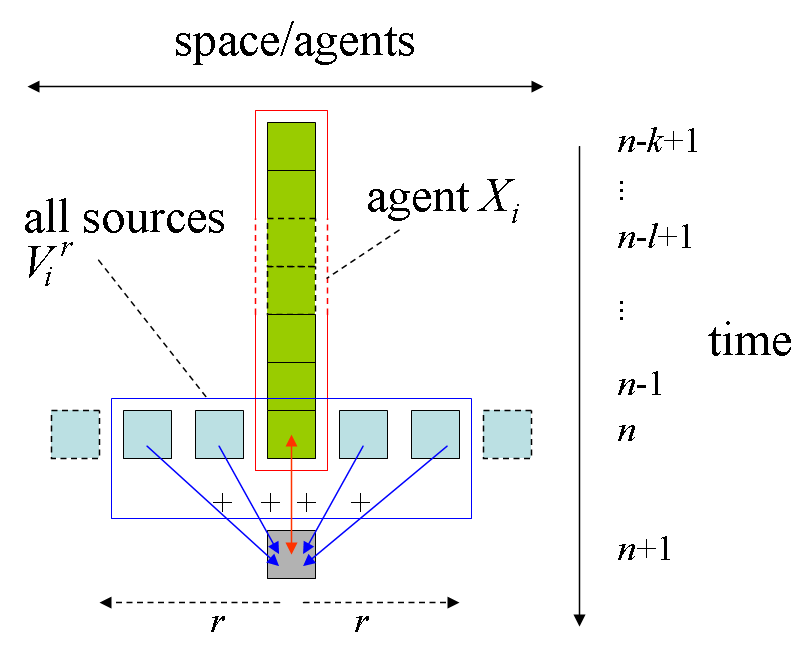}}}
	\caption{\label{fig:transferAndSeperable} 
		  \protect 	\subref{fig:transferEntropy} Transfer Entropy $t(i,j,n+1,k)$: information contained in the source cell $X_{i-j}$ about the next value of the destination cell $X_{i}$ at time $n+1$ in the context of the destination's past.
			 \protect \subref{fig:separableInfo} Separable information $s(i,n+1,k)$: information gained about the next value of the destination from separately examining each causal information source in the context of the destination's past. For CAs these causal sources are within the cell range $r$. (NB: \figu{transferEntropy} is reprinted with kind permission from Springer Science+Business Media: \cite{liz13a} Lizier, J.~T. , {\em {The Local Information Dynamics of Distributed Computation in Complex Systems}}, Springer Theses, Springer, Berlin / Heidelberg, Copyright 2013. \figu{separableInfo} is reprinted with permission from \cite{liz10e} J.~T. Lizier, M.~Prokopenko, and A.~Y. Zomaya, ``Information modification and particle collisions in distributed computation,'' \emph{Chaos},  vol.~20, no.~3, p. 037109, 2010. Copyright 2010, AIP Publishing LLC.)}
\end{figure*}

\citet{schr00} presented \textbf{transfer entropy} as a measure for information transfer in order to address deficiencies in the previous de facto measure, mutual information (\eq{mi}), the use of which he criticized in this context as a symmetric measure of statically shared information.
Transfer entropy is defined as the deviation from independence (in bits) of the state transition of an information destination \textit{X} from the previous state of an information source \textit{Y}:
\begin{equation}
	T_{Y \rightarrow X}(k,l) = \sum_{w_n}
    p(w_n)
    \log_{2}{ \frac{ p(x_{n+1} \mid \vec{x}^{(k)}_{n},\vec{y}^{(l)}_{n})}{p(x_{n+1} \mid \vec{x}^{(k)}_{n})}}
	\label{eq:te},
\end{equation}
where $w_n$ is the state transition tuple $(x_{n+1},\vec{x}^{(k)}_{n},\vec{y}^{(l)}_{n})$.
This is shown diagrammatically in \figu{transferEntropy}.
The transfer entropy will be zero if the next value of the destination is completely dependent on its past (leaving no information for the source to add), or if the state transition of the destination is independent of the destination. At the other extreme, it will be maximal if the state transition is completely specified by the source (in the context of the destination's past).
As such, the transfer entropy is a \textit{directional}, \textit{dynamic} measure of information transfer.
It is a \textit{conditional} mutual information, casting it as the average information in the source about the next state of the destination conditioned on the destination's past.
We have provided a thermodynamic interpretation of transfer entropy in \cite{pro13a}.

The role of the past state of the destination $\vec{x}^{(k)}_{n}$ is particularly important here.
This past state can indirectly influence the next value via the source or other neighbors: this may be mistaken as an independent flow from the source here \cite{liz08a}.
In the context of distributed computation, this is recognizable as the \textit{active information storage}. That is, conditioning on the destination's history $\vec{x}^{(k)}_{n}$ serves to eliminate the active information storage from the transfer entropy measurement. Yet any self-influence transmitted prior to these \textit{k} values will not be eliminated: in \cite{liz08a} we 
suggested that the asymptote $k \rightarrow \infty$ is most correct for variables displaying non-Markovian dynamics. Just as the excess entropy and active information storage require $k \rightarrow \infty$ to capture all information storage, accurate measurement of the transfer entropy requires $k \rightarrow \infty$ to eliminate all information storage from being mistaken as information transfer.
Further to these, even if the destination variable does display Markovian dynamics of order $k$, synergistic interactions between the source and the past of the destination beyond $k$ time steps necessitate the use of a longer destination history to capture the information transfer, again leading us to $k \rightarrow \infty$ to capture all transfer.
We describe other interpretations of the role of $\vec{x}^{(k)}_{n}$ in \cite{liz13c}, including properly capturing the state transition of the destination and capturing the contribution of the source in the context of that state transition; which align with the above.
The most generally correct form of the transfer entropy is therefore computed as:
\begin{equation}
	T_{Y \rightarrow X}(l) = \lim_{k \rightarrow \infty}{\sum_{w_n}
    p(w_n)
    \log_{2}{ \frac{ p(x_{n+1} \mid \vec{x}^{(k)}_{n},\vec{y}^{(l)}_{n})}{p(x_{n+1} \mid \vec{x}^{(k)}_{n})}}}
	\label{eq:teLimit},
\end{equation}
with $T_{Y \rightarrow X}(k,l)$ retained for finite-$k$ estimates.

Also, we note that considering a source \emph{state} $\vec{y}^{(l)}_{n}$ rather than a scalar $y_{n}$ is most appropriate where the observations $y$ mask a hidden causal process in $Y$, or where multiple past values of $Y$ in addition to $y_n$ are causal to $x_{n+1}$.
Otherwise, where $y_n$ is directly causal to $x_{n+1}$, and where it is the only direct causal source in $Y$ (e.g. in CAs), we use only $l=1$ \cite{liz08a,liz10a} and drop it from our notation here.
Furthermore, note that one may use source-destination delays other than one time step, and indeed it is most appropriate to match any causal delay from $Y$ to $X$ \cite{wib13a}.

Next, we introduced the corresponding \textbf{local transfer entropy} at each observation $n$ in \cite{liz08a}:
\begin{align}
	t_{Y \rightarrow X}(n+1,l) = \lim_{k \rightarrow \infty}{t_{Y \rightarrow X}(n+1,k,l)}
	\label{eq:localTE_generic_limit}, \\
	t_{Y \rightarrow X}(n+1,k,l) = \log_{2}{ \frac{ p(x_{n+1} \mid \vec{x}^{(k)}_{n},\vec{y}^{(l)}_{n})}{p(x_{n+1} \mid \vec{x}^{(k)}_{n})}}
	\label{eq:localTE_generic}.
\end{align}
The local transfer entropy describes the information added by a specific source state $\vec{y}^{(l)}_{n}$ about $x_{n+1}$ in the context of the past of the destination $\vec{x}^{(k)}_{n}$.
Of course, we have $T_{Y \rightarrow X}(k,l) = \left\langle t_{Y \rightarrow X}(n+1,k,l) \right\rangle$.

For lattice systems such as CAs with spatially-ordered variables, the local information transfer to agent $X_{i}$ from $X_{i-j}$ (across $j$ cells to the right) at time $n+1$ is represented as:
\begin{align}
	t(i,j,n+1,l) = \lim_{k \rightarrow \infty}{t(i,j,n+1,k,l)}
	\label{eq:localTeLattice_limit}, \\
	t(i,j,n+1,k,l) = \log_{2}{ \frac{ p(x_{i,n+1} \mid \vec{x}^{(k)}_{i,n},\vec{x}^{(l)}_{i-j,n})}{p(x_{i,n+1} \mid \vec{x}^{(k)}_{i,n})}}
	\label{eq:localTeLattice}.
\end{align}
This information transfer $t(i,j,n+1,k,l)$ to variable $X_{i}$ from $X_{i-j}$ at time $n+1$ is illustrated in \figu{transferEntropy}. 
Then $t(i,j,n,k,l)$ is defined for every spatiotemporal destination $(i,n)$, for every information channel or direction $j$; sensible values for $j$ correspond to causal information sources, i.e. for CAs, sources within the cell range $|j| \leq r$.
Again, for homogeneous variables (with stationarity) it is appropriate to estimate the PDFs used in \eq{localTeLattice} from all spatiotemporal observations, and we write the average across homogeneous variables as $T(j,k) = \left\langle t(i,j,n,k) \right\rangle$.

Calculations conditioned on no other information contributors (as in \eq{localTeLattice})) are labeled as \textit{apparent} transfer entropy \cite{liz08a}. Local apparent transfer entropy $t(i,j,n,k)$ may be either positive or negative, with negative values occurring where (given the destination's history) the source element is actually misleading about the next value of the destination.
In deterministic systems, this can only occur where another source is influencing the destination at that time.
To counter that effect, the transfer entropy may be conditioned on other possible causal information sources $Z$, to eliminate their influence from being attributed to the source in question $Y$ \cite{schr00}.
We call this the \textbf{conditional transfer entropy} \cite{liz10e}, given (as a finite-k estimate) along with the \textbf{local conditional transfer entropy} as follows: 
\begin{align}
	T_{Y \rightarrow X \mid Z}(k,l) = \left\langle t_{Y \rightarrow X \mid Z}(n+1,k,l) \right\rangle, \\
	t_{Y \rightarrow X \mid Z}(n+1,k,l) = \log_{2}{ \frac{ p(x_{n+1} \mid \vec{x}^{(k)}_{n},\vec{y}^{(l)}_{n},z_{n})}{p(x_{n+1} \mid \vec{x}^{(k)}_{n},z_{n})}}
	\label{eq:localTE_genericConditional}.
\end{align}
$Z$ may of course be multivariate, or be an embedded state vector $\vec{z}_n^{(m)}$ itself.
Indeed, a special case involves conditioning on \textit{all} sources jointly in the set of causal information contributors $\vec{V}_X$ to $X$, except for the source $Y$,
i.e. $\vec{V}_{X} \setminus Y$.
This gives the \textbf{complete transfer entropy} $T^c_{Y \rightarrow X}(k,l) = T_{Y \rightarrow X \mid \vec{V}_{X} \setminus Y}(k,l)$ \cite{liz08a}.
At time step $n$, this set $\vec{V}_{X} \setminus Y$ has joint state $\vec{v}_{x,y,n}$, 
giving the \textbf{local complete transfer entropy} \cite{liz08a}:\footnote{Note that if past values of $X$ are causal sources to the next value $x_{n+1}$, they can be included in $\vec{v}_{x,y,n}$, but this is irrelevant for complete TE since they are already conditioned on in $\vec{x}_n^{(k)}$.}
\begin{eqnarray}
	t^c_{Y \rightarrow X}(n+1,k,l) = t_{Y \rightarrow X \mid \vec{V}_{X,Y}}(n+1,k,l)
	\label{eq:localTE_genericComplete}, \\
	\vec{v}_{x,y,n} = \left\{ z_n \mid \forall Z \in \vec{V}_{X} \setminus Y \right\}
	\label{eq:neighbourhood_generic}.
\end{eqnarray}
For CAs the set of causal information contributors to $X_i$ is the neighborhood $\vec{V}^r_i$ of $X_i$, and for the complete transfer entropy we condition on this set except for the source $X_{i-j}$: $\vec{V}^r_i \setminus X_{i-j}$.
At time step $n$ this set has joint value $\vec{v}^{r}_{i,j,n}$, giving the following expression for the local complete transfer entropy in CAs \cite{liz08a}:
\begin{eqnarray}
	t^{c}(i,j,n+1,k) = \log_{2}{ \frac
		{p \left( x_{i,n+1} \mid \vec{x}^{(k)}_{i,n},x_{i-j,n},\vec{v}^{r}_{i,j,n} \right)}
		{p \left( x_{i,n+1} \mid \vec{x}^{(k)}_{i,n},\vec{v}^{r}_{i,j,n} \right)}}
	\label{eq:completeTE}, \\
	\vec{v}^{r}_{i,j,n} = \left\{ x_{i+q,n} \mid \forall q: -r \leq q \leq +r, q \neq -j, q \neq 0 \right\}
	\label{eq:neighbourhood}.
\end{eqnarray}
Again, the most correct form is $t^{c}(i,j,n+1)$ in the limit $k \rightarrow \infty$.
In deterministic systems (e.g. CAs), complete conditioning renders $t^{c}(i,j,n) \geq 0$ because the source can only add information about the outcome of the destination. 

\subsection{\label{totalInfoTransfer}Total information, entropy rate and collective information transfer}

The total information required to predict the next value of any process $X$ is the \emph{local entropy} $h_X(n+1)$ \eq{localEntropy}.
Similarly, the \emph{local temporal entropy rate} $h_{\mu X}(n+1,k) = - \log_2{p(x_{n+1} \mid x^{(k)}_n)}$ is the information to predict the next value of that process given that its past, and the entropy rate is the average of these local values: $H_{\mu X}(k) = \left\langle h_{\mu X}(n+1,k) \right\rangle$. For lattice systems we have $h_{\mu}(i,n+1,k)$.
Now, the entropy can be considered as the sum of the active information storage and temporal entropy rate \cite{liz10e,liz12a}:
\begin{align}
		H_{X_{n+1}} = I_{X_{n+1};\vec{X}^{(k)}_n} + H_{X_{n+1} \mid \vec{X}^{(k)}_n}
		\label{eq:entropyMemEntRate1}, \\
		h_X(n+1) = a_X(n+1,k) + h_{\mu X}(n+1,k)
		\label{eq:localEntActEntRate}.
\end{align}

For deterministic systems (e.g. CAs) there is no intrinsic uncertainty, so the local temporal entropy rate is equal to the \textbf{local collective transfer entropy} \cite{liz10a} and represents a collective information transfer: the information about the next value of the destination \textit{jointly} added by the causal information sources in the context of the past of the destination.
This suggested that the local collective transfer entropy (or simply the local temporal entropy rate $h_\mu(i,n,k)$ for deterministic systems) is likely to be a meaningful measure and filter for incoming information.

Also, we showed that the information in a destination variable can be expressed as a sum of incrementally conditioned mutual information terms, considering each of the sources iteratively \cite{liz10e,liz10a}. For ECAs, these expressions become:
\begin{align}
		h(i,n+1) = a(i,n+1,k) + t(i,j=-1,n+1,k) + t^c(i,j=1,n+1,k)
		\label{eq:localInfoEcas},
\end{align}
(and vice-versa in $j=1,-1$).
Clearly, this total information is not simply a simple sum of the active information storage and the apparent transfer entropy from each source, nor the sum of the active information storage and the complete transfer entropy from each source.


\subsection{\label{localTeResults}Local information transfer results}

In this section, we review the application of the local apparent and complete transfer entropies, as well as the local entropy rate, to several ECA rules \cite{liz07b,liz08a,liz10e,liz13a}.
We focus in particular here on the local apparent transfer entropy, whose profiles $t(i,j=1,n,k=16)$ (measuring transfer across one unit to the right per time step) are plotted for rules 54 (\figu{54-te-1-colour-16})
and 22 (\figu{22-te-1-colour-16}), with $t(i,j=-1,n,k=16)$ (transfer across one unit to the left per time step) plotted for rules 54 (\figu{54-te--1-colour-16}), 110 (\figu{110-te--1-colour-16}) and $\phi_{par}$ (\figu{phi-te--1-colour-10}).

Both the local apparent and complete transfer entropy highlight \textbf{particles as strong positive information transfer against background domains}. This is true for both regular gliders as well as domain walls in rule 18 (not shown here, see \cite{liz08a}).
Importantly, the particles are measured as information transfer in their direction of macroscopic motion, as expected.
As such, local transfer entropy provided the first quantitative evidence for the long-held conjecture that particles are the \textit{dominant} information transfer agents in CAs.
For example, at the ``x" marks in \figu{54-closeup} which denote parts of the right-moving $\gamma^+$ gliders, we have $p(x_{i,n+1} \mid \vec{x}^{(k=16)}_{i,n},x_{i-1,n}) = 1.00$ and $p(x_{i,n+1} \mid \vec{x}^{(k=16)}_{i,n}) = 0.25$: there is a strong information transfer of $t(i,j=1,n,k=16) = 2.02$ bits here because the source (in the glider) added a significant amount of information to the destination about the continuation of the glider.

For $\phi_{par}$ we confirm the role of the gliders as information transfer agents in the human understandable computation, and demonstrate information transfer across multiple units of space per unit time step for fast-moving gliders in \figu{phi-te--3-colour-10}.
Interestingly, we also see in \figu{phi-te--1-colour-10} ($j=-1$) and \figu{phi-te--3-colour-10} ($j=-3$) that the apparent transfer entropy can attribute information transfer to several information sources, whereas the complete transfer entropy (see \figu{phi-teComp--1-colour-10}) is more likely to attribute the transfer to the single causal source.
We emphasize though that information transfer and causality are distinct concepts, as discussed in detail in \cite{liz10a}.
This result also underlines that \textbf{the apparent and complete transfer entropies have a similar nature but are complementary} in together determining the next state of the destination (as in \eq{localInfoEcas}). Neither measure is more correct than the other though -- both are required to understand the dynamics fully. A more detailed example contrasting the two is studied for rule 18 in \cite{liz08a}, showing that the complete TE detects transfer to $X$ due to synergies between the source $Y$ and conditioned variable $Z$, whereas the apparent TE does not.

We also examine the profiles of the local temporal entropy rate $h_\mu(i,n+1,k)$ (which is equal to the local collective transfer entropy in these deterministic systems) here in 
\figu{110-entRate-colour-16} for rule 110
and \figu{22-entRate-colour-16} for rule 22.
As expected, the local temporal entropy rate profiles $h_\mu(i,n+1,k)$ highlight particles moving in \emph{each} relevant channel and are a useful \emph{single} spatiotemporal filter for moving emergent structure.
In fact, these profiles are quite similar to the profiles of the negative values of local active information storage. This is not surprising given they are counterparts in \eq{localEntActEntRate}: where $h_\mu(i,n+1,k)$ is strongly positive (i.e. greater than 1 bit), it is likely that $a(i,n+1,k)$ is negative since the local single cell entropy will average close to 1 bit for these examples. Unlike $a(i,n+1,k)$ however, the local temporal entropy rate $h_\mu(i,n+1,k)$ is never negative.

Note that while achieving the limit $k \rightarrow \infty$ is not computationally feasible, a large enough $k$ was required to achieve a reasonable estimates of the transfer entropy; without this, as discussed earlier the active information storage was not eliminated from the transfer entropy measurements in the domains, and the measure did not distinguish the particles from the domains \cite{liz08a}.

We also demonstrated \cite{liz08a} that while there is zero information transfer in an infinite periodic domain (since the dynamics there only involve information storage), there is a small non-zero information transfer in domains acting as a background to gliders, effectively indicating the \emph{absence} of gliders. These small non-zero information transfers are stronger in the wake of a glider, indicating the absence of (relatively common) following gliders. Similarly, we note here that the local temporal entropy rate profiles $h_\mu(i,n+1,k)$ contain small but non-zero values in these periodic domains.
Furthermore, there is interesting structure to the information transfer in the domain of rule 18, described in detail in \cite{liz08a}.
As such, while particles are the dominant information transfer agents in CAs, they are not the only transfer entities.

The highlighting of structure by local transfer entropy is similar to results from other methods of filtering for structure in CAs \cite{sha06,wue99,han92,hel04}, but subtly different in revealing the leading edges of gliders as the major transfer elements in the glider structures, and providing multiple profiles (one for each direction or channel of information transfer). 

Also, a particularly relevant result for our purposes is the finding of negative values of transfer entropy for some space-time points in particles moving orthogonal to the direction of measurement in space-time. This is displayed for $t(i,j=1,n,k=16)$ in rule 54 (\figu{54-te-1-colour-16}), and $t(i,j=-1,n,k=16)$ for rule 110 (\figu{110-te--1-colour-16}), and also occurs for rule 18 (see \cite{liz08a,liz13a}). In general this is because the source, as part of the domain, suggests that this same domain found in the past of the destination is likely to continue; however since the next value of the destination forms part of the particle, this suggestion proves to be misinformative.
For example, consider the ``x'' marks in \figu{54-closeup} which denote parts of the right-moving $\gamma^+$ gliders. If we now examine the source at the right (still in the domain), we have $p(x_{i,n+1} \mid \vec{x}^{(k=16)}_{i,n},x_{i+1,n}) = 0.13$, with $p(x_{i,n+1} \mid \vec{x}^{(k=16)}_{i,n}) = 0.25$ as before, giving $t(i,j=-1,n,k=16) = -0.90$ bits: this is negative because the source (still in the domain) was misinformative about the destination.

Regarding the local information transfer structure of rule 22, we note similar results as for local information storage. There is much information transfer here (in fact the average value $T(j=1,k=16) = 0.19$ bits is greater than for rule 110 at 0.07 bits), although there is no coherent structure to this transfer. Again, this demonstrates the utility of local information measures in providing more detailed insights into system dynamics than their global averages.

In this section, we have described how the local transfer entropy quantifies the information transfer at space-time points within a system, and provides evidence that particles are the dominant information transfer agents in CAs.
We also described the collective transfer entropy, which quantifies the joint information contribution from all causal information contributors, and in deterministic systems is equal to the temporal entropy rate. However, we have not yet separately identified collision events in CAs: to complete our exploration of the information dynamics of computation, we now consider the nature of information modification.


\section{\label{modification}Information Modification}

\citet{lang90} interpreted information modification as interactions between transmitted and/or stored information which resulted in a modification of one or the other. CAs provide an illustrative example, where the term \textit{interactions} is generally interpreted to mean collisions of particles (including blinkers as information storage), with the resulting dynamics involving something other than the incoming particles continuing unperturbed. The resulting dynamics could involve zero or more particles (with an annihilation leaving only a background domain), and perhaps even some of the incoming particles.
Given the focus on perturbations in the definition here, it is logical to associate a collision event with the modification of transmitted and/or stored information, and to see it as an information processing or decision event. Indeed, as an information processing event the important role of collisions in determining the dynamics of the system is widely acknowledged \cite{hord01}, e.g. in the $\phi_{par}$ density classification.

Attempts have previously been made to quantify information modification or processing in a system \cite{san02,yam94,kin06a}. However, these have either been too specific to allow portability across system types (e.g. by focusing on the capability of a system to solve a known problem, or measuring properties related to the particular type of system being examined), focus on general processing as movement or interpretation of information rather than specifically the modification of information, or are not amenable to measuring information modification at \textit{local} space-time points \textit{within} a distributed system.

In this section, we review the separable information \cite{liz10e} as a tool to detect non-trivial information modification events, and demonstrate it as the first measure which filtered collisions in CAs as such.
At the end of the section however, we describe criticisms of the separable information, and describe current efforts to develop new measures of information modification.

\subsection{\label{sepInfo}Local separable information}

We begin by considering what it means for a particle to be \textit{modified}. For the simple case of a glider, a modification is simply an alteration to the predictable periodic pattern of the glider's dynamics. At such points, an observer would be surprised or misinformed about the next value of the glider, having not taken account of the entity about to perturb it.
The intuition behind the separable information \cite{liz10e} is that this interpretation is reminiscent of the earlier findings that local apparent transfer entropy $t(i,j,n)$ and local active information storage $a(i,n)$ were negative where the respective information sources were \textit{misinformative} about the next value of the information destination (in the context of the destination's past for transfer entropy).
Local active information storage was misinformative at gliders, and local apparent transfer entropy was misinformative at gliders traveling in the orthogonal direction to the measurement in space-time.
This being said, one expects that the local apparent transfer entropy measured in the direction of glider motion will be \textit{more informative} about its evolution than any misinformation conveyed from other sources.
However, where the glider is modified by a collision with another glider, we would no longer expect the local apparent transfer entropy in its macroscopic direction of motion to remain informative about the dynamics. Assuming that the incident glider is also perturbed, the local apparent transfer entropy in its macroscopic direction of motion will also not be informative about the dynamics at this collision point. We expect the same argument to be true for irregular particles, or domain walls.


As such, we made the hypothesis that at the spatiotemporal location of a local information modification event or collision, \textit{separate} inspection of each information source will \textit{misinform} an observer overall about the next value of the modified information destination. More specifically, the information sources referred to here are the past history of the destination (via the local active information storage) and each other causal information contributor (examined in the context of the past history of the destination, via their local apparent transfer entropies).

We quantified the independent sum of information gained from separate observation of the information storage and information transfer contributors $Y \in V$ to a process $X$ as the \textbf{local separable information} $s_X(n)$ \cite{liz10e}:
\begin{eqnarray}
	s_X(n) = a_X(n) + \sum_{Y \in V, Y \neq X}{ t_{Y \rightarrow X}(n) }
	\label{eq:separableInfo}.
\end{eqnarray}
$s_X(n,K)$ is used for finite-$k$ estimates.
For CAs, where the causal information contributors are homogeneously within the neighborhood $r$, we write the local separable information in lattice notation as:
\begin{eqnarray}
	s(i,n) = a(i,n) + \sum_{j=-r,j\neq0}^{+r}{ t(i,j,n) }
	\label{eq:seperableInfoCAs}. 
\end{eqnarray}
We use $s(i,n,k)$ to represent finite-$k$ estimates, and show $s(i,n,k)$ in \figu{separableInfo}.

As inferred earlier, we expected the local separable information to be \textit{positive} or \textit{highly separable} where separate observations of the information contributors are informative overall regarding the next value of the destination. This was be interpreted as a trivial information modification, because information storage and transfer are not interacting in any significant manner. More importantly, we expected the local separable information to be \textit{negative} at spatiotemporal points where an information modification event or collision takes place. Here, separate observations are misleading overall because a \textit{non-trivial information modification} is taking place (i.e. the information storage and transfer are interacting).

Importantly, this formulation of non-trivial information modification aligns with the descriptions of complex systems as consisting of (a large number of) elements interacting in a \textit{non-trivial} fashion \cite{pro09}, and of emergence as where \textit{``the whole is greater than the sum of its parts''}.
``The whole'' meant to refer to examining all information sources together; the whole is greater where all information sources must be examined together in order to receive positive information on the next value of the examined entity. 
The thinking behind the separable information was in the direction of measuring synergies between information storage and transfer sources, prior to the development of a proper framework for examining such synergies \cite{will10a}, as discussed in \secRef{modificationOutlook}.

\subsection{\label{sepInfoResults}Local separable information results}

Next, we review the application of the separable information to several ECA rules from \cite{liz10e}.
The simple gliders in ECA rule 54 give rise to relatively simple collisions which we focus on in our discussion here.
Notice that the positive values of $s(i,n,k=16)$ for rule 54 (displayed in \figu{54-sep-colour-16}) are concentrated in the domain regions and at the stationary gliders ($\alpha$ and $\beta$). As expected, these regions are undertaking trivial computations only.
The negative values of $s(i,n,k=16)$ are also displayed in \figu{54-sep-colour-16}, with their positions marked.
The dominant negative values are clearly concentrated around the areas of collisions between the gliders, including collisions between the traveling gliders only (marked by ``A'') and between the traveling gliders and the stationary gliders (marked by ``B'' and ``C'').

Collision ``A'' involves the $\gamma^+$ and $\gamma^-$ particles interacting to produce a $\beta$ particle ($\gamma^+ + \gamma^- \rightarrow \beta$ \cite{hord01}). The only information modification point highlighted is one time step below (or delayed from) that at which the gliders initially appear to collide (see close-up of raw values in \figu{54-closeup}).
The periodic pattern in the past of the destination breaks there, however the neighboring sources are still able to support separate prediction of the value (i.e. $a(i,n,k=16) = -1.09$ bits, $t(i,j=1,n,k=16) = 2.02$ bits and $t(i,j=-1,n,k=16) = 2.02$ bits, giving $s(i,n,k=16) = 2.95$ bits). This is no longer the case however where our measure has successfully identified the modification point; there we have $a(i,n,k=16) = -3.00$ bits, $t(i,j=1,n,k=16) = 0.91$ bits and $t(i,j=-1,n,k=16) = 0.90$ bits, with $s(i,n,k=16) = -1.19$ bits suggesting a non-trivial information modification.
A delay is also observed before the identified information modification points of collision types ``B'' and ``C''; 
possibly these delays represent a time-lag of information processing. Not surprisingly, the results for these other collision types imply that the information modification points are associated with the creation of new behavior: in ``B'' and ``C'' these occur along the newly created $\gamma$ gliders, and for ``C'' 
in the new $\alpha$ blinkers.

Importantly, weaker non-trivial information modification points continue to be identified at every second point along all the $\gamma^+$ and $\gamma^-$ particles after the initial collisions.
These can also be seen for a similar (right-moving) glider in rule 110 in \figu{110-sep-colour-16}).
This was unexpected from our earlier hypothesis. However, these events can be understood as non-trivial computations of the continuation of the glider in the \textit{absence} of a collision; in effect they are virtual collisions between the real glider and the absence of an incident glider.
Interestingly, this finding is analogous to the small but non-zero information transfer in periodic domains indicating the absence of gliders.

We also note that measurements of local separable information must be performed with a reasonably large value of $k$.
Here, using $k < 4$ could not distinguish any information modification points clearly from the domains and particles, and even $k < 8$ could not distinguish \textit{all} the modification points (results not shown).
Correct quantification of information modification requires satisfactory estimates of information storage and transfer, and accurate distinction between the two.

We observe similar results in $s(i,n,k=10)$ for $\phi_{par}$ (see \figu{phi-sep-colour-10}).
Note that the collisions at the left and right of the figure do in fact contain significant negative values of $s(i,n,k=10)$ -- around 1 to 2 bits -- however these are difficult to see in comparison to the much larger negative value at the collision in the centre of the diagram. 
These results confirm the particle collisions here as non-trivial information modification events, and this therefore completes the evidence for all of the conjectures about this human understandable computation.

The results for $s(i,n,k=16)$ for ECA rule 110 (see \figu{110-sep-colour-16}) are also similar to those for rule 54.
Here, we have collisions ``A'' and ``B'' which show non-trivial information modification points slightly delayed from the collision in a similar fashion to those for rule 54. We note that collisions between some of the more complex glider structures in rule 110 (not shown) exhibit non-trivial information modification points which are more difficult to interpret, and which are even more delayed from the initiation of the collision.
The larger delay is perhaps this is a reflection of the more complex gliders requiring more time steps for the processing to take place.
An interesting result not seen for rule 54 is a collision where an incident glider is absorbed by a blinker, without any modification to the absorbing glider (not shown here, see \cite{liz10e}). No information modification is detected for this absorption event by $s(i,n,k=16)$: this is as expected because the information storage for the absorbing blinker is sufficient to predict the dynamics at this interaction.

As a further test of the measure, we examined collisions between the domain walls of rule 18; see \cite{liz10e}. We found that collisions between the domain walls were quite clearly highlighted as the dominant information modification events for this rule - importantly, this result provides evidence that collision of \emph{irregular} particles are information modification events, as expected.
The reader is referred to \cite{liz10e} for further discussion of the information modification dynamics of rule 18.

We also apply $s(i,n,k=16)$ to ECA rule 22, as displayed in \figu{22-sep-colour-16}. As could be expected from our earlier results, there are many points of both positive and negative local separable information here. The presence of negative values implies the occurrence of non-trivial information modification, yet there does not appear to be any structure to these profiles. Again, this aligns well with the lack of coherent structure found using the other measures in this framework and from the local statistical complexity profile of rule 22 \cite{sha06}.

\subsection{\label{sec:modificationOutlook}Outlook for information modification}

Here, we have reviewed the local separable information, which attempts to quantify information modification at each spatiotemporal point in a complex system.
The separable information suggests that information modification events occur where the separable information is negative, indicating that separate or independent inspection of the causal information sources (in the context of the destination's past) is misleading because of non-trivial interaction between these sources.
The local separable information was demonstrated to provide the first quantitative evidence that \textbf{particle collisions in CAs are the dominant information modification events} therein,
and is capable of identifying events involving both creation and destruction.

With that said however, it has been shown that the separable information double-counts parts of the information in the next state of the destination \cite{fleck11a}.
This is clear, and so it is a heuristic more than a measure.
Efforts to properly quantitatively define information modification, by combining information dynamics with the partial information decomposition approach \cite{will10a} to properly measure synergies between information storage and transfer, are ongoing and described in \cite{liz13b}.
While the separable information is not a proper information-theoretic measure, it remains the only technique which has uniquely filtered particle collision events.

\section{\label{coherence}Importance of coherent computation}
Our framework has proven successful in locally identifying the component operations of distributed computation. We then considered in \cite{liz10d} whether this framework can provide any insights into the overall complexity of computation. In other words, what can our results say about the difference in the complex computations of rules 110 and 54 as compared to rule 22 and others? We review those considerations in this section.

We observed that the \textit{coherence} of local computational structure appears to be the most significant differentiator here. ``Coherence" implies a property of sticking together or a logical relationship \cite{oed}: in this context we use the term to describe a logical spatiotemporal relationship between values in local information dynamics profiles. For example, the manner in which particles give rise to similar values of local transfer entropy amongst spatiotemporal neighbors is coherent.
From the spatiotemporal profiles presented here, we note that rules 54 and 110 exhibit the largest amount of coherent computational structure, with rule 18 containing a smaller amount of less coherent structure. Rules 22 and 30 (results for rule 30 not shown, see \cite{liz10d}) certainly exhibit all of the elementary functions of computation, but do not appear to contain any coherent structure to their computations.
This aligns well with similar explorations of local information structure for these rules, e.g. by \citet{sha06}.
Using language reminiscent of Langton's analysis \cite{lang90}, we suggested that complex systems exhibit very \textit{highly-structured coherent} computation in comparison to ordered systems (which exhibit coherence but minimal structure in a computation dominated by information storage) and chaotic systems (whose computations are dominated by rampant information transfer eroding any coherence).

\begin{figure*}[t]
	\subfloat[110]{\fbox{\label{fig:110-stateSpace-teRightvsActive}\includegraphics[width=0.49\textwidth]{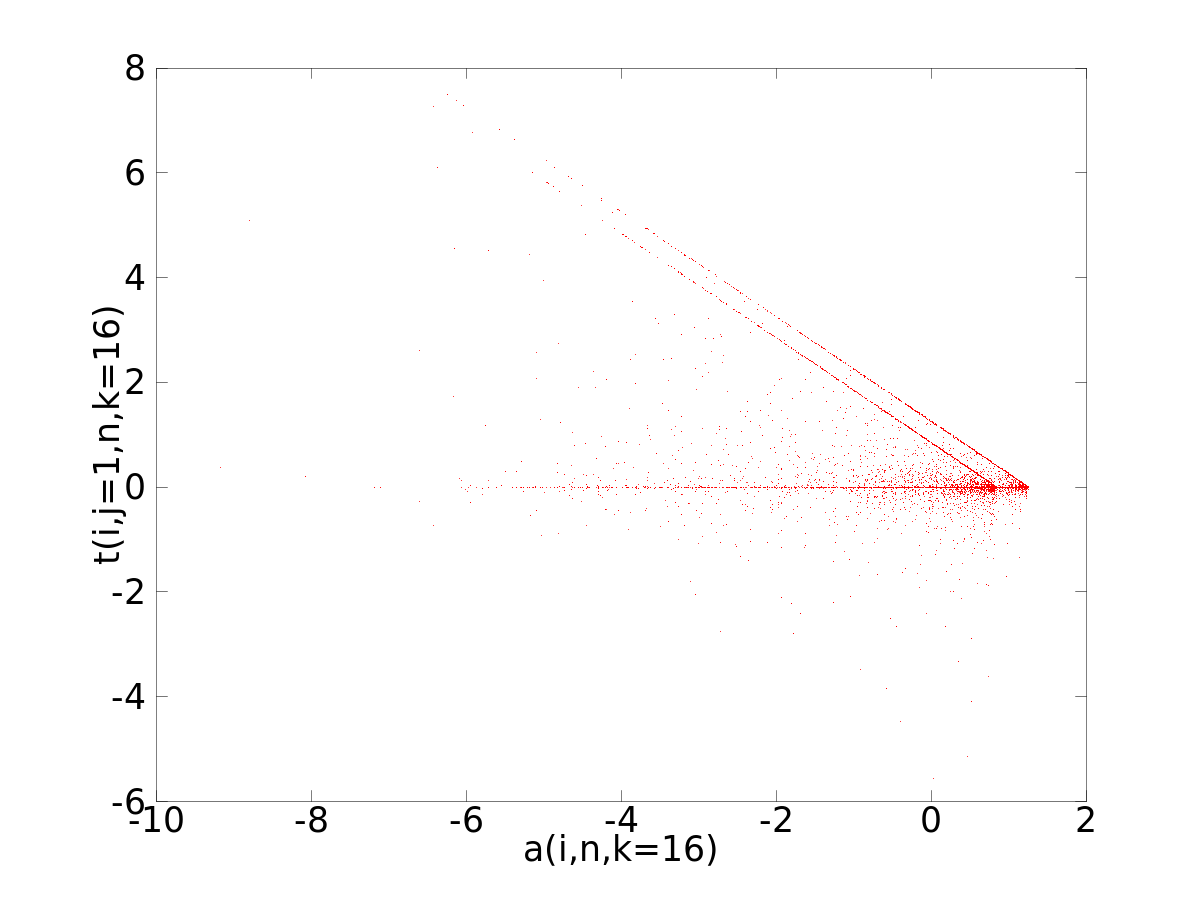}}}
	\subfloat[54]{\fbox{\label{fig:54-stateSpace-teRightvsActive}\includegraphics[width=0.49\textwidth]{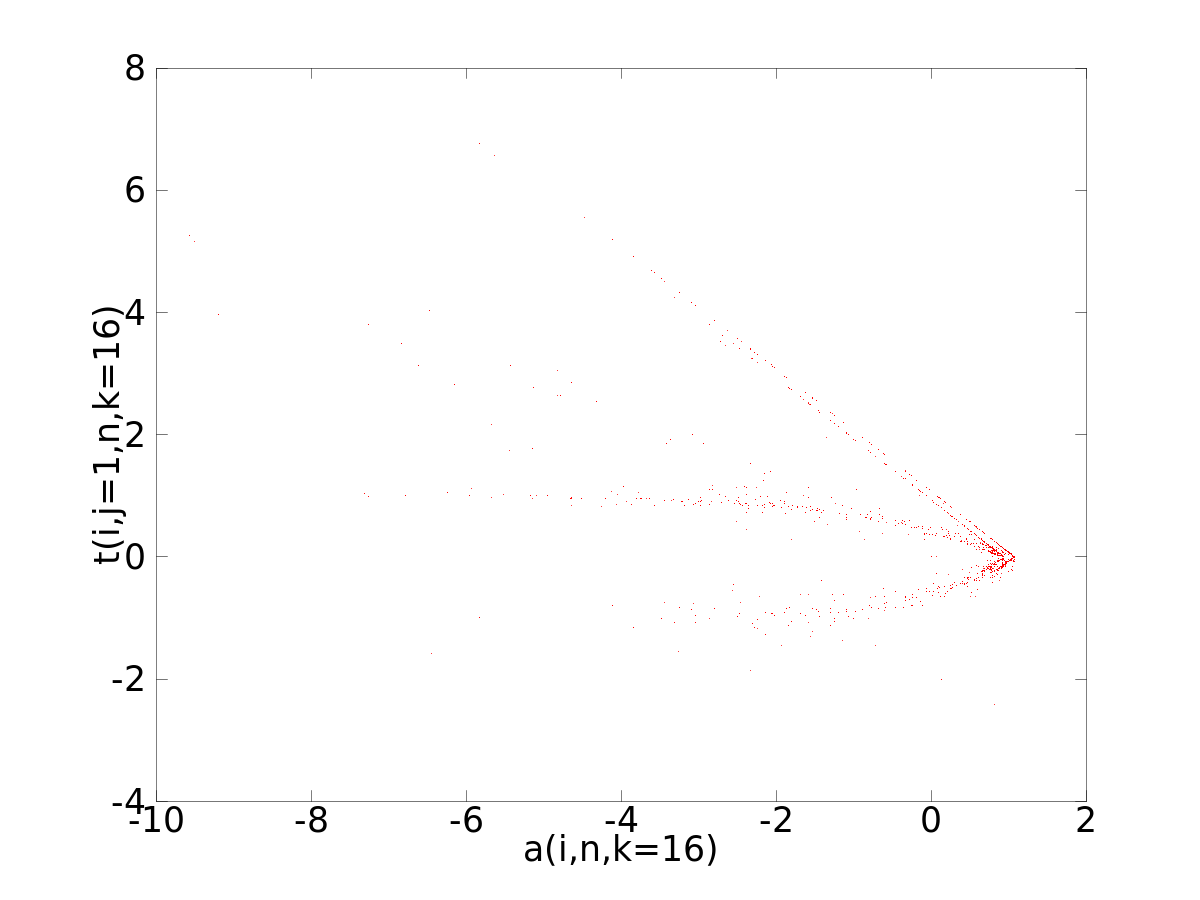}}}
	
	\subfloat[30]{\fbox{\label{fig:30-stateSpace-teRightvsActive}\includegraphics[width=0.49\textwidth]{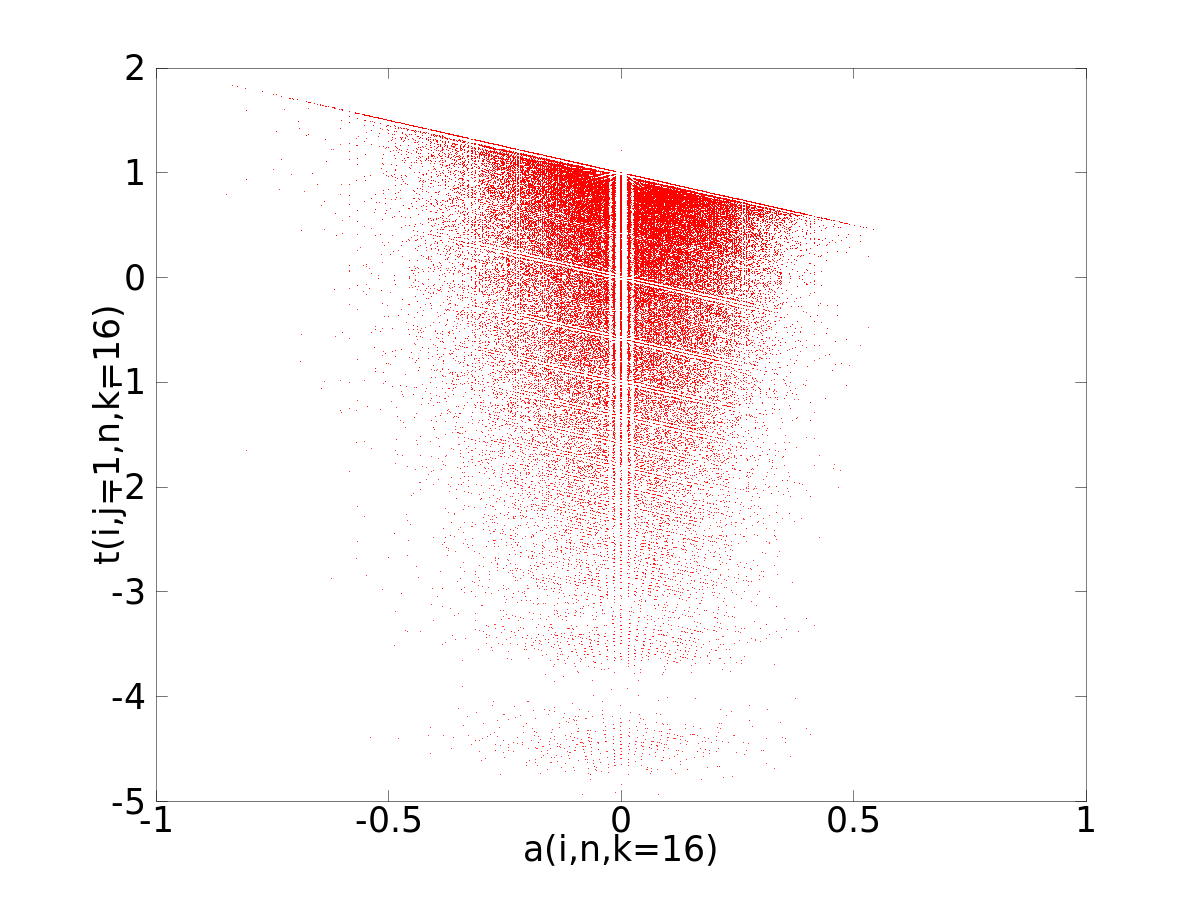}}}
	\subfloat[22]{\fbox{\label{fig:22-stateSpace-teRightvsActive}\includegraphics[width=0.49\textwidth]{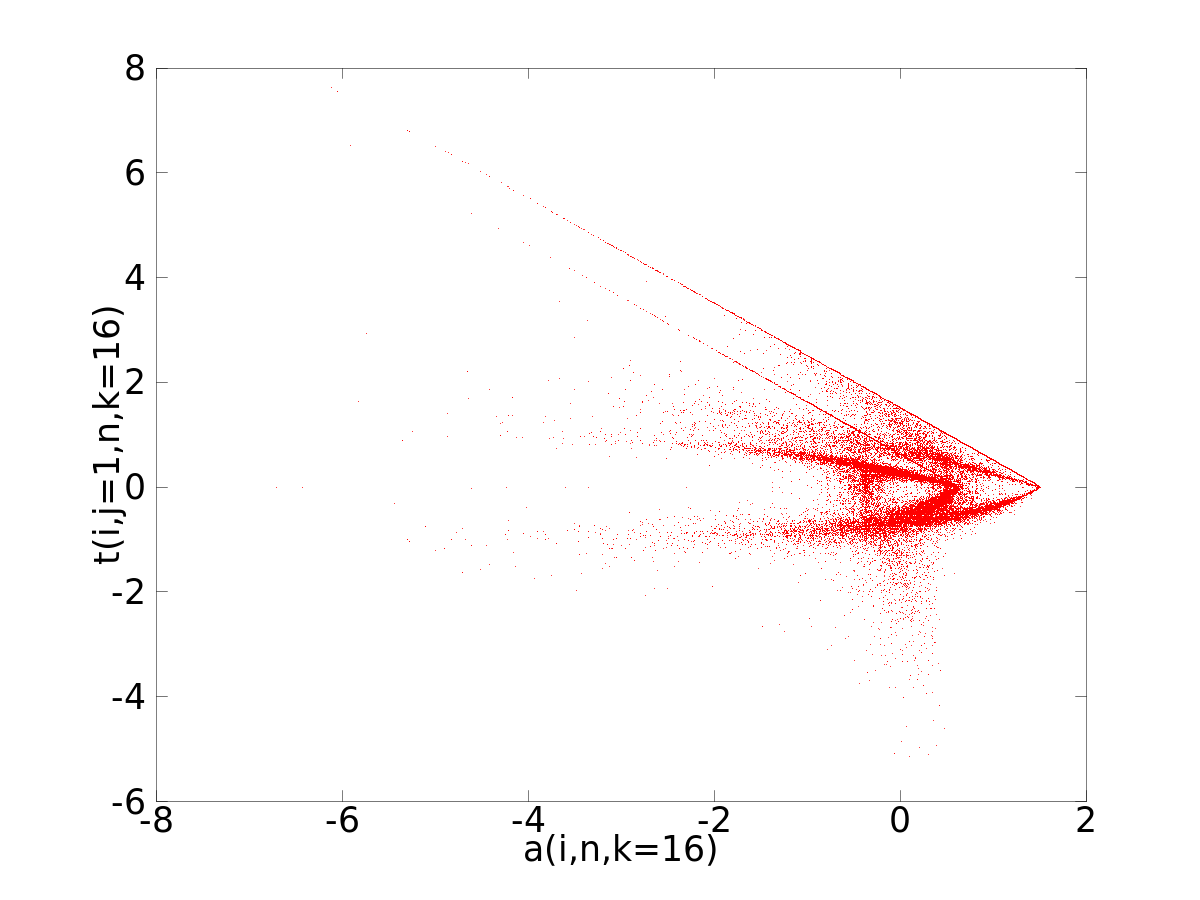}}}
	\caption{\label{fig:stateSpaceDiagrams} State space diagrams of local transfer entropy (one step to the right) $t(i,j=1,n,k=16)$ versus local active information $a(i,n,k=16)$ (both in bits) at the same space-time point $(i,n)$ for several ECA rules:  \protect \subref{fig:110-stateSpace-teRightvsActive} 110,  \protect \subref{fig:54-stateSpace-teRightvsActive} 54,  \protect \subref{fig:30-stateSpace-teRightvsActive} 30 and  \protect \subref{fig:22-stateSpace-teRightvsActive} 22 (after \citep{liz10d}).
	}
\end{figure*}

Coherence may also be interpreted as a logical relationship \textit{between} profiles of the individual local information dynamics (as three axes of complexity) rather than only within them. To investigate this possibility, \figu{stateSpaceDiagrams} plots state-space diagrams of the local apparent transfer entropy for $j=1$ versus local active information storage (after \cite{liz10d}). Each point in these diagrams represents the local values of each measure at one spatiotemporal point, thereby generating a complete state-space for the CA.
Such state-space diagrams are known to provide insights into structure that are not visible when examining either measure in isolation; for example, in examining structure in classes of systems (such as logistic maps), \citet{feld08} demonstrate that plotting average excess entropy versus entropy rate (while changing a system parameter) reveals loci of the two which are not clear from observing either in isolation. Here however we are looking at structure \textit{within} a single system rather than across a class of systems.

The state-space diagram for rule 110 (\figu{110-stateSpace-teRightvsActive}) exhibits interesting structure, with significant clustering around certain areas and lines in the state space, reflecting its status as a complex rule. (The two diagonal lines are upper limits representing the boundary condition $t^c(i,j=-1,n,k=16) \geq 0$ for both destination states ``0'' and ``1'').
Rule 54 (\figu{54-stateSpace-teRightvsActive}) exhibits similar structure in its state-space diagram.
On the other hand, the example state space diagram for rule 30 (\figu{30-stateSpace-teRightvsActive}) exhibits minimal structure (apart from the mathematical upper limit), with a smooth spread of points across the space reflecting its underlying chaotic nature.
From the apparent absence of coherent structure in its space-time information profiles, one may expect state-space diagrams for rule 22 to exhibit a similar absence of structure to rule 30. As shown by \figu{22-stateSpace-teRightvsActive} however this is not the case: the state-space diagram for rule 22 exhibits significant structure, with similar clustering to that of rules 110 and 54.

Importantly, the apparent information structure in the state-space diagrams lends some credence to the claims of complex behavior for rule 22 discussed in Section \ref{caExamples}. However it is a very subtle type of structure, not complex enough to be revealed in the individual local information profiles shown here or by other authors (e.g. by \citet{sha06}). The structure does not appear to be coherent in these individual profiles, though the state space diagrams indicate a coherent relationship between the local information dynamics which may underpin coherent computation at other scales.

There are certain clues as to the type of coherence which may be displayed by rule 22. \figu{22-te-1-colour-16} does appear to have some traces of coherent transfer entities  moving diagonally in space-time; however these seem to be distributed through the CA, seemingly without structure or interactions.
More concretely, \citet{grass83b} observed that for rule 22, ``there are (at least) four different sets of ordered states, corresponding to $S_i(t)=0$ for all even/odd $i$ and all even/odd $t$'' -- i.e. \textit{rule 22 does have a domain pattern which self-replicates} (with four possible configurations, just offset from each other in space and time). Indeed, $\epsilon$-machines have been generated to recognize these domains \cite{crutch13a}. \citet{grass83b} goes on to note that ``In contrast to the ordered states of rule 18, these states however are unstable: after implanting a kink in an otherwise ordered state, the kink widens without limit, leaving behind it a seemingly disordered state.'' That is to say, \textit{this domain pattern does not self-organise and it is not robust to perturbations}. 
This means that, despite the existence of such domains, they are highly unlikely to be found ``in the wild'' (i.e. when rule 22 is started from random initial states, as we have done for \figu{22}).
One could also view this as inferring that ``life in one dimension'' (the perspective that rule 22 is a 1D projection of the 2D Game of Life \cite{mcin90}) is less stable than ``life in two dimensions''.

Coming back to \figu{22-stateSpace-teRightvsActive} -- it is possible that these domain patterns, or small versions of them, are what is detected as a signature of coherent information structure by our methods above.
Furthermore, emerging evidence suggests that rule 22 can be set up in certain initial states which sustain such domains for a longer period, with certain stable domain walls \cite{crutchEccs}, and that these domain walls are detected as information transfer by our methods.
Our investigations in this area remain ongoing.

Given the subtlety of structure in the bounds of our analysis, and using our mutual information heuristics, at this stage we conclude that the behavior of this rule is less complex than that exhibited by rules 110 and 54.
As such, we suggested that coherent information structure is a defining feature of complex computation, and explored a technique for inferring this property using local information dynamics. These state-space diagrams for local information dynamics produced useful visual results and were shown to provide interesting insight into the nature of computation in rule 22.

\section{\label{conclusion}Conclusion}

In this chapter, we have reviewed our complete quantitative framework for the information dynamics of distributed computation in complex systems. Our framework quantifies the information dynamics in terms of the component operations of universal computation: information storage, information transfer and information modification.
Our framework places particular importance on examining computation on a local scale in space and time. While averaged or system-wide measures have their place in providing summarized results, this focus on the local scale is vital for understanding the information dynamics of computation and provides many insights that averaged measures cannot.

We reviewed the application of the framework to cellular automata, an important example because of the weight of previous studies on the nature of distributed computation in these systems. Significantly, our framework provided the first quantitative evidence for the widely accepted conjectures that blinkers provide information storage in CAs, particles are the dominant information transfer agents, and particle collisions are the dominant information modification events. In particular, this was demonstrated for the human-understandable density classification computation carried out by the rule $\phi_{par}$. This is a fundamental contribution to our understanding of the nature of distributed computation, and provides impetus for the framework to be used for the analysis and design of other complex systems.

The application to CAs aligned well with other methods of filtering for complex structure in CAs. However, our work is distinct in that it provides several different views of the system corresponding to each type of computational structure.
In particular, the results align well with the insights of computational mechanics, underlining the strong connection between these approaches.

From our results, we also observed that coherent local information structure is a defining feature of complex distributed computation, and used local information state-spaces to study coherent complex computation. 
Here, our framework provides further insight into the nature of computation in rule 22 with respect to the accepted complex rules 54 and 110.
Certainly rule 22 exhibits all of the elementary functions of computation, yet (in line with \citet{sha06}) there is no apparent coherent structure to the profiles of its local information dynamics (``in the wild'' at least). On the other hand, state space views of the interplay between these local information dynamics reveal otherwise hidden structure. Our framework is unique in its ability to resolve both of these aspects. We conclude that rule 22 exhibits more structure than chaotic rules, yet the subtlety of this structure prevents it from being considered as complex as rules 110 and 54.

The major thrust of our work since the presentation of this framework was to apply it to other systems, because the information-theoretic basis of this framework makes it readily applicable as such.
For example, we have used the measures in this framework to: quantitatively demonstrate coherent waves of motion in flocks and swarms as information cascades \cite{wang12a}; evolve a modular robot for maximal information transfer between components, observing the emergence of glider-like information cascades \cite{liz08d}; and to study interactions in robotic football and the relation of information measures to success on the field \cite{cliff13a}.
We have also inferred information structure supporting cognitive tasks using fMRI brain imaging data \cite{liz11a}, and studied how the computational capabilities of artificial neural networks relate to underlying parameters and ability to solve particular tasks \cite{boed12a}.
We have also made more specific investigations of the relationship between underlying network structure and computational capabilities, including: revealing that intrinsic information storage and transfer capabilities are maximized near the phase transition in dynamics for random Boolean networks \cite{liz08c}; showing that regular networks are generally associated with information storage, random networks with information transfer, and small-world networks exhibit a balance of the two \cite{liz11b}; revealing that feedback and feedforward loop motifs determine information storage capability \cite{liz12b}; and exploring how these information measures relate to synchronization capability of network structures \cite{ceg11}.
We have also explored the relationship of the framework to the context of the observer \cite{liz13c}, and provided thermodynamic interpretations of transfer entropy \cite{pro13a} and related information-theoreic quantities \cite{pro11a}.
And finally, we have begun reformulating our approach to information modification in seeking a proper measure rather than a heuristic \cite{liz13b}.
Further developments in all of these directions are expected in the future, due to the utility of the framework.

\section*{Acknowledgements}
The authors thank Melanie Mitchell for helpful comments and suggestions regarding an early version of this manuscript.



\backmatter
\printindex


\end{document}